\documentclass{jfm}
\usepackage{graphicx,subfigure}
\usepackage{natbib}
\usepackage{color}

\usepackage[usenames,dvipsnames,svgnames,table]{xcolor}
\definecolor{darkgreen}{rgb}{0,0.5,0}

\ifCUPmtlplainloaded \else
  \checkfont{eurm10}
  \iffontfound
    \IfFileExists{upmath.sty}
      {\typeout{^^JFound AMS Euler Roman fonts on the system,
                   using the 'upmath' package.^^J}%
       \usepackage{upmath}}
      {\typeout{^^JFound AMS Euler Roman fonts on the system, but you
                   dont seem to have the}%
       \typeout{'upmath' package installed. JFM.cls can take advantage
                 of these fonts,^^Jif you use 'upmath' package.^^J}%
      }
  \else
  \fi
\fi

\ifCUPmtlplainloaded \else
  \checkfont{msam10}
  \iffontfound
    \IfFileExists{amssymb.sty}
      {\typeout{^^JFound AMS Symbol fonts on the system, using the
                'amssymb' package.^^J}%
       \usepackage{amssymb}%
         
         \let\geq=\geqslant
      }{}
  \fi
\fi

\ifCUPmtlplainloaded \else
  \IfFileExists{amsbsy.sty}
    {\typeout{^^JFound the 'amsbsy' package on the system, using it.^^J}%
     \usepackage{amsbsy}}
    {\providecommand\boldsymbol[1]{\mbox{\boldmath $##1$}}}
\fi

\newsavebox{\astrutbox}
\sbox{\astrutbox}{\rule[-5pt]{0pt}{20pt}}

\usepackage[retainorgcmds]{IEEEtrantools}

\title[Low-Reynolds number swimming in a capillary tube]{Low-Reynolds number swimming in a capillary tube}

\author[Lailai Zhu, Eric Lauga and Luca Brandt]%
{L. Zhu$^1$%
  \thanks{Email address for correspondence: lailai@mech.kth.se},\ns
E. Lauga$^2$ 
and L. Brandt$^1$}

\affiliation{$^1$Linn\'e Flow Centre, KTH Mechanics, S-100 44 Stockholm, Sweden\\[\affilskip]
$^2$Department of Mechanical and Aerospace Engineering, University of California San Diego, 9500 Gilman Drive, La Jolla CA 92093-0411, USA}

\pubyear{2010}
\volume{650}
\pagerange{119--126}
\date{}

\begin{document}
\maketitle 

\begin{abstract}
We use the boundary element method to study the low-Reynolds number locomotion of a spherical model microorganism in a circular tube. The swimmer propels itself by tangential or normal surface motion in a tube whose radius is on the order of the swimmer size. Hydrodynamic interactions with the tube walls significantly affect the average swimming speed and power consumption of the model microorganism. In the case of swimming parallel to the tube axis, the locomotion speed is always reduced (resp.~increased) for swimmers with tangential (resp.~normal) deformation. In all cases, the rate of work necessary for swimming is increased by confinement. Swimmers with no force-dipoles in the far field generally follow helical trajectories, solely induced by hydrodynamic interactions with the tube walls, and  in qualitative agreement with recent experimental observations for {\textit{Paramecium}}. Swimmers of the puller type always display stable locomotion at a location which depends on the strength of their force dipoles: swimmers with weak dipoles (small $\alpha$) swim in the centre of the tube while those with  strong dipoles (large $\alpha$) swim near the walls. In contrast, pusher swimmers and those   employing normal deformation are unstable and end up crashing into the walls of the tube. Similar dynamics is observed for swimming into a curved tube. These results could be relevant for the future design of artificial microswimmers in confined geometries. 
 \end{abstract}


\section{Introduction}

The locomotion of self-propelled microorganisms have recently attracted sizable attention in both the applied mathematics and biophysics communities~\citep{lighthill_book,lighthill76,brennen77,purcell77,yates_review,bergreview00,fauci06,LaugaSwim}. A number of novel phenomena have been discovered, including the dancing behaviour of pair \textit{Volvox} algae~\citep{dauncevolvex}, the collective motion of motile
\textit{Bacillus subtilis} bacteria~\citep{dombrowski04}, and tumbling dynamics of flagellated \textit{Chlamydomonas}~\citep{tumbleStealth_Polin,tumbleStealth_Roman}. 
One area of particularly active research addresses the variation in  cell mobility as a response to complex environments,  including the dependence on  the  rheological properties of the medium where cells swim ~\citep{lauga07,Fu08,elfring_syn,LiuHelixPnas,PRL_Shen_Visco,lailai-pre,laipof1}, the presence of an external shear flow~\citep{ecoli_upstream_yale,shear_ecoli_kaya}, gravity~\citep{gryotaxis_stocker}, or a sudden aggression~\citep{pnas_hamel}.

Many microorganisms swim close to boundaries, and as a result the effect of boundaries on fluid-based locomotion has been extensively studied. \textit{E.~coli} bacteria display circular trajectories near boundaries, clockwise when the wall is rigid~\citep{eric06_biophy_circle} and anti-clockwise near a free surface~\citep{swimImage}. Experiments, simulations, and theoretical analysis are employed to investigate locomotion near a plane wall~\citep{katz_sperm_bound, katz_slend_bound, ramia93,fauci_95_sperm,goto05,ericPRL_08,spermSmithJfm,shumHeadhelix, Saverio_farfield_jfm} explaining in particular the accumulation of cells by boundaries~\citep{ramia93, fauci_95_sperm,ericPRL_08,spermSmithJfm, shumHeadhelix,gold_swim_noise}. 
Most of these past studies consider the role of hydrodynamic interaction in the kinematics and energetics of micro-scale locomotion, developing fundamental understanding of how
microorganisms swim in confined geometries. 

Although most past studies consider interactions with a single planar, infinite surface,  microorganisms in nature are faced with  more complex geometries. For example, mammalian spermatozoa  are required to swim through narrow channel-like
passages~\citep{Winet_pipe73,katz_sperm_bound}, \textit{Trypanosoma} protozoa  move in narrow blood vessels~\citep{Winet_pipe73},  and bacteria often have to navigate  microporous environments such as soil-covered
beaches and river-bed sediments~\citep{bac_porous}. 

Locomotion of microorganisms in strongly confined geometries is therefore biologically relevant, and a few studies have been devoted to its study. An experimental investigation was conducted by \cite{Winet_pipe73} to measure the wall drag on ciliates freely swimming in a tube. Perturbation theory was employed to analyse the swimming speed and efficiency of an infinitely long model cell swimming along the axis of a tube~\citep{Felderhol_pof_pipe}. {Numerical simulations using  multiple-particle collision dynamics were carried out to study the motion of model microswimmers in a cylindrical Poiseuille flow~\citep{swim_pois}.} Recent experiments~\citep{sunny_para_tube}, which originally inspired the present paper, showed that \textit{Paramecium} cells tend to follow helical trajectories when self-propelling inside a capillary tube. 

In this article, we model the locomotion of ciliated microorganisms inside a capillary tube. Specifically, we  develop a boundary element method (BEM) implementation of the locomotion of the squirmer model ~\citep{modelIntro,Blake1971a} inside straight and curved capillary tubes. The boundary element method has been successfully used in the past to simulate self-propelled cell locomotion at low Reynolds numbers~\citep{ramia93,swimInter-Pedley,shumHeadhelix,wavingring}. Our specific computational approach is tuned to deal with strong geometrical confinement whereas traditional BEM show inaccuracy when the tube becomes too narrow~\citep{poz_tube2005}.

After introducing the mathematical model, its computational implementation and validation, we calculate the swimming speed and power consumption of spherical squirmers with different swimming gaits inside a straight or curved capillary tube. The effect of tube confinement, swimming gait,  and cell position is investigated.  By  studying trajectories of squirmers with varying initial cell positions and orientations, we show that cells end up either swimming parallel to the tube axis or performing wavelike motions with increasing/decreasing wave magnitudes. The dynamic stability of the cell motion is also analysed  revealing the importance of the swimming gaits. In particular, squirmers employing the gait leading to minimum work against the surrounding fluid  are seen to generically execute helical trajectories, in agreement with the experimental observation of swimming
\textit{Paramecia} inside a capillary tube~\citep{sunny_para_tube}.

\section{Mathematical Model}\label{mathmodel} 

\subsection{Squirmer model}
In this work we use steady squirming as a  model for the locomotion of ciliated cells such as \textit{Paramecium}  -- more specifically, as a model for the envelope of the deforming cilia tips at the surface of the cells. This steady model has been employed in the past to address fundamental processes in  the physics of swimming microorganisms, such as nutrient uptake~\citep{nutrientPedleysteady}, locomotion in stratified and viscoelastic fluids~\citep{swimPycnoclines,laipof1}, biomixing~\citep{biostirring}, and the collective behaviour of microorganisms~\citep{PhysRevLett.100.088103,graham_suspension,Art_squirmer}. Furthermore, simulations of two interacting \textit{Paramecium} using the squirmer model showed good agreement with corresponding experiments~\citep{Ishikawa_TwoProlate}.

In the model, a non-zero velocity, $\mathbf{u}_{ST}$, is imposed at the surface of the spherical swimmer 
 as first proposed by  \cite{modelIntro} and \cite{Blake1971a}. In this work, we consider for the most part pure tangential surface deformation (normal surface deformation will be covered in \S \ref{jet} only) and adopt the concise formulation introduced in~\citet[]{PhysRevLett.100.088103} where the imposed velocity  on the surface of a squirmer centred at the origin is explicitly given as
\begin{equation}
 \mathbf{u}_{ST} ({\bf r})=\sum_{n\geq1} \frac{2}{n(n+1)}B_n
  P_{n}'\left(\frac{\hat{\bf{e}} \cdot \mathbf{r}}{r}\right)
  \left( \frac{\hat{\mathbf{e}} \cdot \mathbf{r}}{r} \frac{\mathbf{r}}{r}-\hat{\mathbf{e}} \right),
\end{equation}
where $\hat{\mathbf{e}}$ is the orientation vector of the squirmer, $B_n$ is the $n$th mode of the
tangential surface squirming velocity \citep{Blake1971a}, $P_{n}$ and $P^{\prime}_{n}$ are the $n$th Legendre polynomial and its derivative with respect to the argument,
$\mathbf{r}$ is the position vector, and $r=|\mathbf{r}|$. 
In a Newtonian fluid, the swimming speed of the squirmer in free space is $U_{ST}^{F}=2B_{1}/3$ \citep{Blake1971a} and
thus  dictated by the
first mode only. The second mode, $B_2$, governs the signature of the flow field in the far field (stresslet). As in many previous studies~\citep{swimInter-Pedley,PhysRevLett.100.088103}, we assume $B_{n}=0$ for
$n>2$. In that case, the  power consumption by the swimmer is $\mathcal{P}_{ST}^{F}={8}\pi \mu a \left(2B_{1}^2 + B_{2}^2 \right)/{3}$, where $\mu$ is the dynamic viscosity of the fluid and $a$  the radius of the sphere. 

The tangential velocity on the sphere in the co-moving frame is therefore simply expressed, in spherical coordinates,  as
$u_{\theta}(\theta)=B_{1}\sin\theta+(B_{2}/2 )\sin2\theta$, where $\theta=\arccos (\hat{\mathbf{e}} \cdot
\mathbf{r}/r )$ is the polar angle between the position vector $\mathbf{r}$ and the swimming direction $\hat{\mathbf{e}}$.  We introduce an additional dimensionless parameter, $\alpha$, representing the ratio of the second to the first squirming mode, $\alpha =B_{2}/B_{1}$. When $\alpha$ is positive, the swimmer is called a puller and obtain the impetus from its front part. As $\alpha$ is negative, the cell is called a pusher and thrust is generated from the rear of the body. A puller (resp.~pusher) generates jet-like flow away from  (resp.~towards) its sides, as shown in~\citet[]{Ishikawa_interface} and references therein. A squirmer with $\alpha=0$ is termed a neutral squirmer, and it is associated with a potential velocity field. 

We note that the  model we consider does not capture the  unsteadiness of the flow arising from the periodic beating of flagella and cilia in  microorganisms such as  \textit{Paramecium} or \textit{Volvox} \citep{JeffPrlOsci,DrescherPrl}. Here we assume that the steady, time-averaged, velocity  dominates the overall dynamics, and will consider the  underlying unsteadiness in future work.
\subsection{Swimming in a tube}
\begin{figure}
   \centering
   \includegraphics[width=0.4 \textwidth]{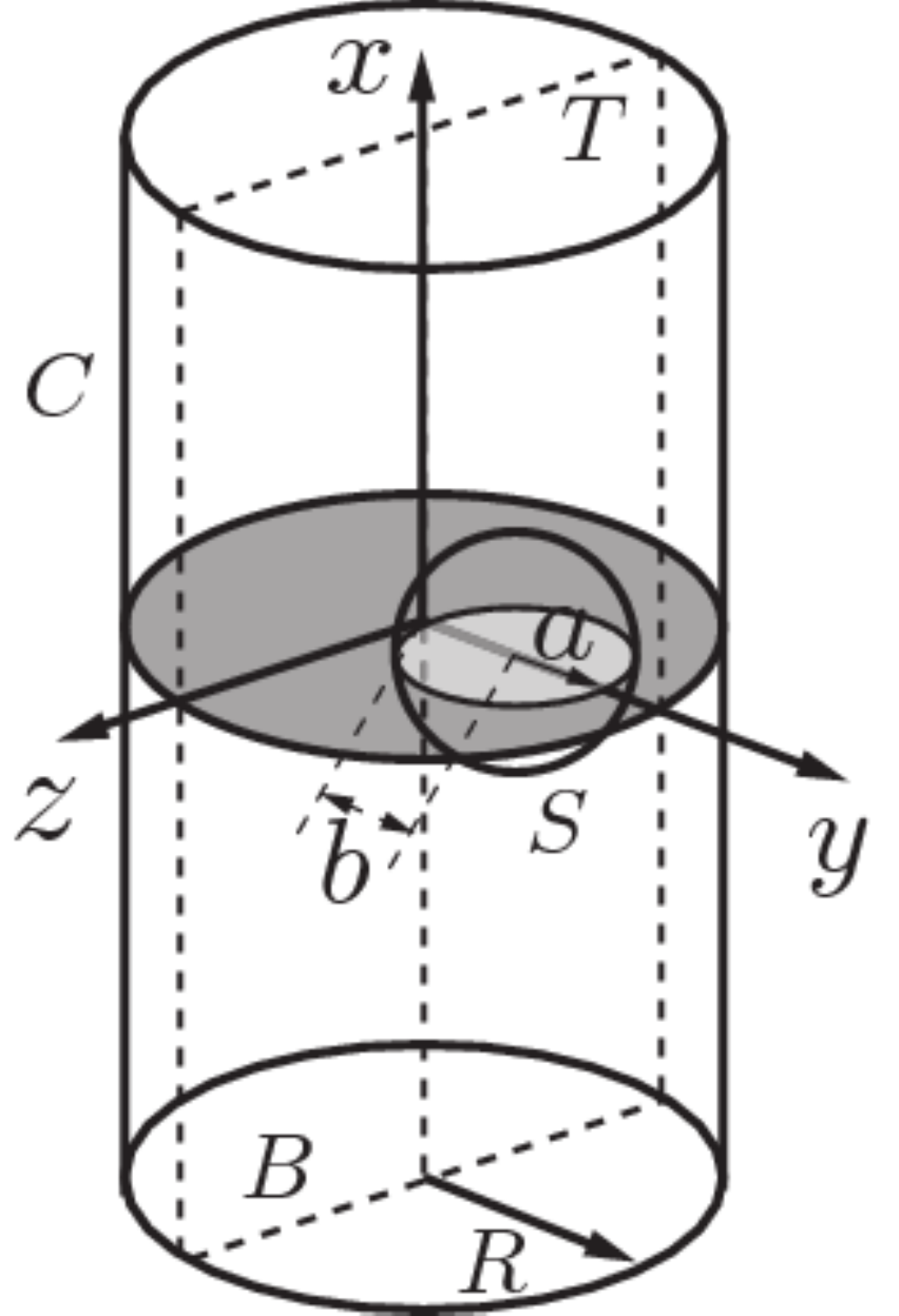} 
   \caption{Schematic representation of a spherical squirmer of radius $a$ swimming in a tube of radius $R$. The centre of the squirmer is located at a distance $b$ from the tube axis.  The origin of the Cartesian coordinates coincides with
the centre of the tube. The bounding surfaces to the fluid are denoted  $S$ (surface of squirmer), $B$ (bottom tube cap), $T$ (top tube ca,p) and $C$ (surface of the tube conduit).}
   \label{fig:tube_sketch}
\end{figure}

The spherical squirmer (radius, $a$) is swimming in a cylindrical tube of radius $R$, as illustrated in figure~\ref{fig:tube_sketch}. The centre of the squirmer is located at a distance $b$ from the tube axis. We use Cartesian coordinates  with an origin at the centre of the tube and the $x$-direction  along the tube axis. As in~\citet[]{tube_higdon} we introduce the nondimensional position
$\beta$ as 
\begin{equation}\label{betadef}
\beta = b/(R-a),
\end{equation}
so that $\beta=0$ indicates that the squirmer is at the centre of the tube while for $\beta=1$ the
squirmer is in perfect contact with the tube wall.

\section{Numerical method}
\subsection{Formulation}
The boundary element method (BEM) has already been successfully adopted to study the hydrodynamics of swimming microorganisms in the Stokesian regime~\citep{ramia93,swimInter-Pedley,shumHeadhelix}. Our current work mainly follows the approach in \citet[]{poz_blue}, the important difference being that we use quadrilateral elements instead of triangle elements as typically used and originally proposed. The
method is introduced briefly here. 

In the Stokesian realm, fluid motion is governed by the Stokes equation
\begin{equation}\label{equ:stokes}
-\nabla p + \mu\nabla^{2}\mathbf{u}=0,
\end{equation}
where $p$ is the dynamic pressure and $\mathbf{u}$  the fluid velocity. Due to the linearity of the Stokes equation,
the velocity field, $\mathbf{u}\left(\mathbf{x}\right)$, resulting from moving bodies with smooth boundary
$S$ can be expressed as
 \begin{equation}{\label{equ:stokeslet}}
 \mathbf{u}\left(\mathbf{x}\right) = \frac{1}{8\pi\mu}\int_{S}\mathbf{f}\left(\mathbf{x}'\right) \cdot
\mathbf{S}\left(\mathbf{x},\mathbf{x}' \right)dS_{\mathbf{x}'},
\end{equation}
where $\mathbf{f}\left(\mathbf{x}'\right)$ is the unknown force per unit area exerted by the body onto the fluid. The tensor $\mathbf{S}$ is the Stokeslet Green's function
 \begin{equation}
 \mathbf{S}_{ij}\left(\mathbf{x},\mathbf{x}'\right) = \left(\frac{\delta_{ij}}{d} +
\frac{d_{i}d_{j}}{d^{3}}\right),
\end{equation}
with $d_{i}=x_{i}-x'_{i}$, $d^{2}=|\mathbf{x}-\mathbf{x}'|^{2} = d_{1}^2+d_{2}^2+d_{3}^2$, and 
$\delta_{ij}$
denoting the Kronecker delta tensor.

We discretize the two bodies in the problem, namely the spherical squirmer and the surrounding tube, into $N$ zero-order elements with  centres at the locations $\{\mathbf{x}_{q},\,q=1\to N\}$, with $q=1\to N_{S}$ denoting the elements on the squirmer surface  and $q=N_{S}+1\to N$ the elements on the surface of the tube.
For the $r^{th}$ element, $\mathbf{f}\left(\mathbf{x}'\right)$ is assumed to be constant over the element and is thus approximated by the value $\mathbf{f}_{r}$. As a consequence, the discretized version of ~\ref{equ:stokeslet} is, when evaluated on one of the elements, 
\begin{equation}{\label{equ:dis_stokeslet}}
\mathbf{u}\left(\mathbf{x}_{q}\right)=\frac{1}{8\pi\mu}\sum_{r=1}^{N}{\mathbf{f}_{r}} \cdot \int_{S_{r}}
\mathbf{S}\left(\mathbf{x}_{q},\mathbf{x}' \right)dS_{\mathbf{x}'},\,q=1\to N.
\end{equation}
In its discrete form, equation~\ref{equ:dis_stokeslet} represents a total of $3N$ equations  for the $3N$ unknown force density components. 

\subsection{Swimming and squirmer boundary conditions}

On the squirmer surface, the left-hand side of \ref{equ:dis_stokeslet} is not fully known. The swimmer has an instantaneous surface deformation, $\mathbf{u}_{S}$, plus $6$ unknown components, namely 
its  instantaneous  translational velocity vector, 
$\mathbf{U}$,  and its instantaneous rotational velocity vector, $\boldsymbol{\Omega}$. Thus, the left hand side  of  \ref{equ:dis_stokeslet}, when evaluated on the surface of the squirmer,  becomes $\mathbf{u}\left(\mathbf{x}_{q}\right)=\mathbf{U}+\boldsymbol{\Omega}\times\widetilde{\mathbf{x}}_{q}+\mathbf{u}_{S}\left(\mathbf{x}_{q}\right)$ for $q$ from
$1$ to $N_{S}$ (here $\widetilde{\mathbf{x}}_{q}=\mathbf{x}-\mathbf{x}^{R}$, where $\mathbf{x}^{R}$ is an arbitrary
reference point, the centre of the spherical squirmer for convenience). The  $6$ additional equations necessary to close the linear system are the force- and torque-free swimming conditions, namely
\begin{eqnarray} \label{eq:balance}
 \int \mathbf{f}\left(\mathbf{x}\right) dS_{\mathbf{x}}  =0 ,\,
 \int \widetilde{\mathbf{x}}\times\mathbf{f}\left(\mathbf{x}\right) dS_{\mathbf{x}}  =0,
\end{eqnarray}
for $\mathbf{x} \in \mathrm{squirmer}$.

\subsection{Other boundary conditions}
The situation addressed in our paper is that of  a squirmer swimming inside an infinitely long tube filled with a quiescent fluid. Numerically, we close both ends of the tube with appropriate  boundary conditions. If the tube caps are sufficiently far away from the squirmer, the velocity near the caps is almost zero, so we have $\mathbf{u}_{B}=\mathbf{u}_{T}=\mathbf{0}$, and the pressure over the bottom and top cap is $p_{B}$ and $p_{T}$ respectively~\citep{poz_tube2005}. The force density $\mathbf{f}_{T}$ over the top
 cap can be approximated by $\mathbf{f}_{T} = p_{T}\mathbf{n}$~\citep{poz_tube2005}, where $\mathbf{n}$
is the unit normal vector pointing from the top cap into the fluid domain. Since pressure is defined up to an arbitrary constant, without loss of generality, we set $p_{T}=0$, $\mathbf{f}_{T} = \mathbf{0}$ and the top cap does not requires  discretization. However, unlike \cite{poz_tube2005}, we do perform discretization on the bottom cap, solving for the normal and tangential components of the force density $\mathbf{f}_{B}$
there.  For the conduit part of the tube, we use no-slip boundary condition, thus write $\mathbf{u}_{C}=0$. 

Since we set the velocity on both caps of the tube to be zero, the error due to domain truncation need to be carefully
considered.  {A truncated tube length $L$ of $\pi R$ or $2\pi R$ was chosen in~\citet[]{poz_tube2005} and $L=3R$ in~\citet[]{tube_higdon}. In our computation of hydrodynamic force on a moving sphere inside, we tested different values $L$ and examined the truncation error. We find the length, $L=2\pi R$, to be long enough for required accuracy (see figure~\ref{fig:relaErr_sph_pipe} and details below). In the case of swimming squirmers, we set $L=3\pi R$, and larger values of $L$ were shown to have  negligible differences in the results.}

\subsection{Discretization and integration}\label{subs:mesh}
Zero-order constant quadrilateral elements are used to discretize all the surfaces. We use six-patch structured grid to
discretize the sphere \citep{tube_higdon,cortezFauci_regusto, djsmith_royal}, mapping six faces of a cube onto the
surfaces of a sphere with each face latticed into a square mesh. The
conduit part of the tube is divided into cylindrical quadrilateral elements obtained from the intersections of evenly spaced planes normal to tube axis and evenly spaced azimuthal planes \citep{tube_higdon, poz_tube2005,sphcell_tube_wen}.
Moreover, orange-like quadrilateral elements are used for the bottom cap of the tube  \citep{tube_higdon,
sphcell_tube_wen}. For the sphere we adopt the six-patch quadrilateral grid with parameterized coordinates instead of triangle elements \citep{poz_blue,poz_tube2005}. Such  discretization with its natural
parametrisation facilitates  Gauss-Legendre quadrature when performing numerical integration. Template points used in the quadrature lie exactly on the sphere surface since their coordinates are derived from the parametrisation. The resulting improved quadrature gives superior accuracy (see Table~\ref{table:sph_unbound}). The integration for singular elements are performed by using plane polar coordinates with Gauss-Legendre quadrature~\citep{poz_blue}.

\begin{figure}
   \centering
   \includegraphics[width=0.6 \textwidth]{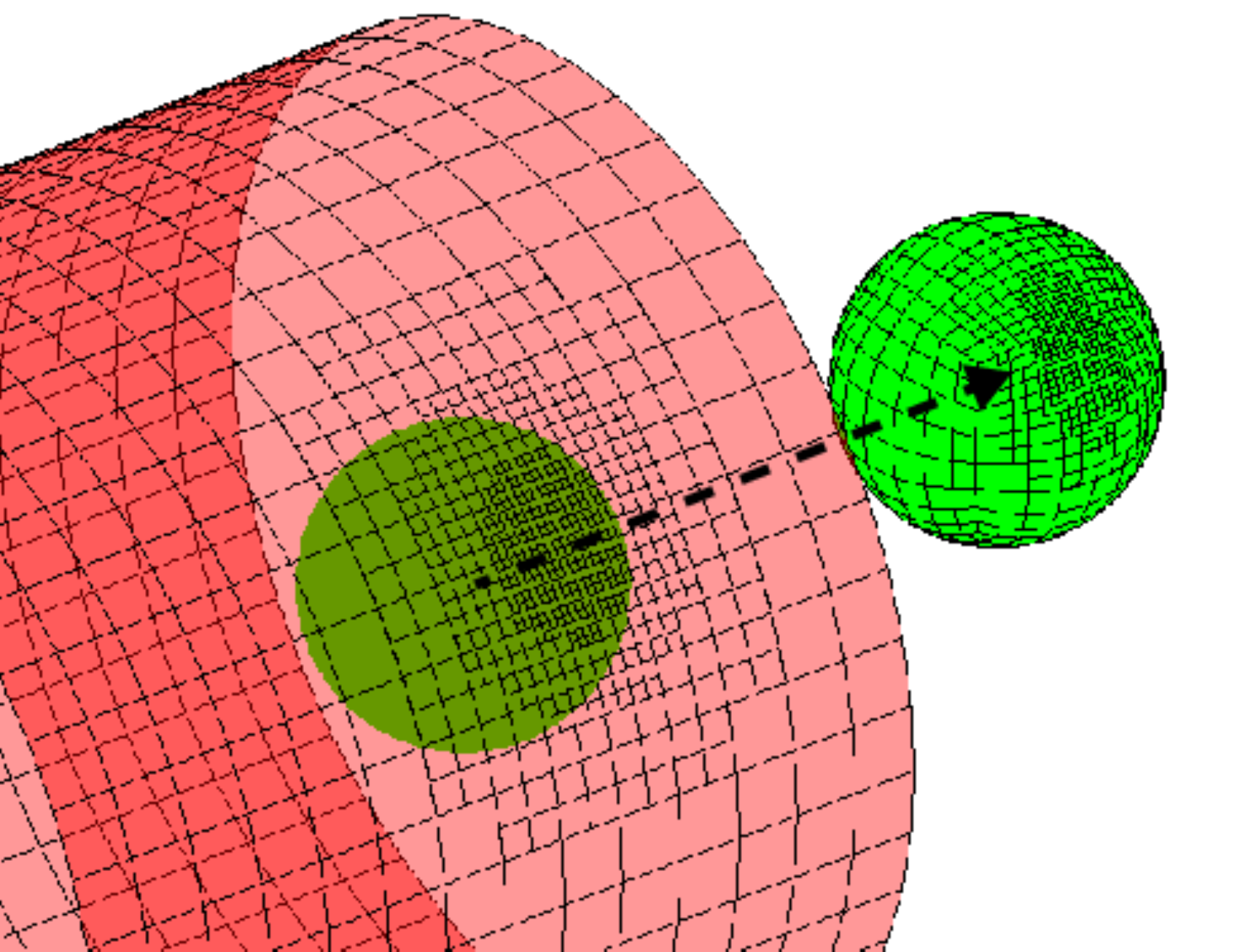} 
   \caption{Local mesh refinement of the cylinder (red) and the sphere (green). The geometrical parameters are $a/R=0.3$ and $\beta=0.95$. For a better visualisation, the mesh on the squirmer surface is reproduced  on the displaced sphere as indicated by the black arrow.}
   \label{fig:mesh}
\end{figure}

In many instances, the squirmer is so close to the cylindrical wall that near-singular integration has to
be performed, a key point to achieve the required accuracy and efficiency~\citep{curse_nearsing}. We
perform local mesh refinement in the near-contact regions
between the squirmer and the tube \citep{swimInter-Pedley,Ishikawa_TwoProlate} as illustrated in figure~\ref{fig:mesh}. The agreement between numerical results with our method and existing results from high-order spectral boundary element method~\citep{tube_higdon} improves significantly when applying such local mesh refinement as shown in the next section where we compute the resistance of a translating sphere inside a cylindrical tube. 

\subsection{Validation and accuracy}~\label{subs:valid}
\begin{figure}
\centering
\includegraphics[width=0.6 \textwidth]{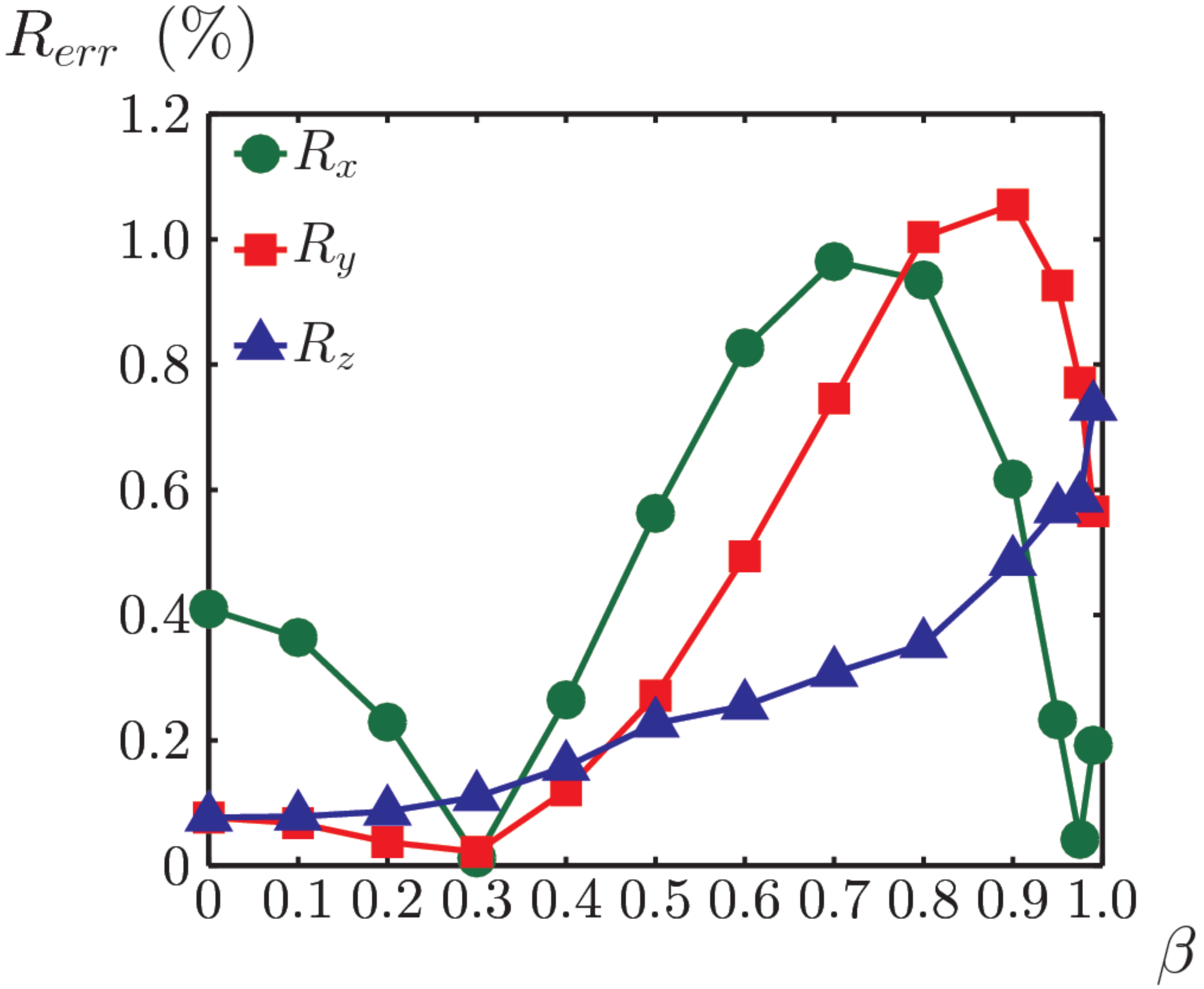} 
\caption{Relative error in the three components of the drag force, $R_x$, $R_y$, and $R_z$,  
on a sphere translating inside a tube ($a/R=0.4$) between the present paper and  \cite{tube_higdon}. Note that the three largest values of $\beta$ chosen are $0.95$, $0.975$ and $0.99$.}
\label{fig:relaErr_sph_pipe}
\end{figure}

We first compute the drag force, $F$, on a translating sphere in an unbounded domain and compare it with the analytical
expression, $F=6\pi\mu a U$, where $\mu$ is the dynamic viscosity of the fluid and $U$ is the translational speed of the sphere. As shown in
Table~\ref{table:sph_unbound}, the current method is very accurate when compared to the three similar approaches~\citep{poz_blue,cortezFauci_regusto,djsmith_royal}. We then compute the drag force and torque on a sphere translating parallel to an infinite, flat, no-slip surface. The  surface is modelled by a discretized plate of size $40a \times 40a$. Our simulation agree well with analytical results~\citep{sph_near_wall}, as shown in Table~\ref{table:sph_near_wall}. Finally, we
compute the drag force acting on a sphere translating inside the tube with confinement $a/R=0.4$, up to a maximum value of $\beta=0.99$,  and compare our results with published data obtained with high-order spectral boundary element method~\citep{tube_higdon}. As illustrated in figure~\ref{fig:relaErr_sph_pipe}, the maximum relative error is less than $1.2\%$. In all simulations, the maximum  confinement is taken to be $\beta = 0.99$ to ensure sufficient accuracy.

\begin{table}
\begin{center}
\begin{tabular}{lcccc}
 & ~\cite{cortezFauci_regusto} & ~\cite{djsmith_royal} & ~\cite{poz_blue} & \quad This paper \quad\\ [3pt]
Element Order  &  0                  & 0                 & 0                & 0 \\ 
(functional variation)             &                     &                   &                  &  \\ 
Element Type   & Quad                & Quad              & Tri              & Quad \\ 
Element Number & $6\times12\times12$ & $6\times6\times6$ & 512              & $6\times6\times6$ \\ 
Singular       & \quad Regularization \quad      & \quad Regularization \quad   & Analytical       & Analytical\\ 
integration    & $\epsilon=0.01$     & $\epsilon=0.01$   & integration      & integration                 \\ 
               &                     & with adaptive     & with             & with                 \\
               &                     & Gauss  & Gauss & Gauss        \\
               &                     & \quad Quadrature \quad & \quad Quadrature \quad & \quad Quadrature \quad       \\ 
Relative error ($\%$)& 12.6 & 0.431 & 9.6\,$10^{-3}$ & 1.4\,$10^{-5}$ \\
\end{tabular}
\caption{\label{table:sph_unbound}
Relative error, in percentage, in the drag force on a translating sphere in an unbounded domain between the method in this paper and  three other methods. The parameter $\epsilon$ is the regularization parameter first introduced in \cite{cortezFauci_regusto}.}
\end{center}
\end{table}

\begin{table}
\begin{center}
\begin{tabular}{ccc}
 $h/a$ & $F_{err}\left(\%\right)$  & $T_{err}\left(\%\right)$ \\ [3pt]
3.7622 & 0.00426 & 0.09488  \\ 
2.3523 & 0.01911 & 0.37879  \\ 
1.5431 & 0.04274 & 0.20478  \\ 
1.1276 & 0.07809 & 0.25773  \\ 
1.0453 & 0.09405 & 0.74217  \\ 
~~~1.005004 & 0.17669 & 1.13493 \\
~~~1.003202 & 0.27472 & 1.74313 \\
\end{tabular}
\caption{\label{table:sph_near_wall}
Relative error in the drag force, $F_{err}$,  and torque,  $T_{err}$, in percentage,  on a sphere translating parallel to an infinite wall between our computations and the  analytical results~\citep{sph_near_wall}. Here $a$ is the radius of the sphere and $h$  the distance between the centre of the sphere and the wall.}
\end{center}
\end{table}

\section{Swimming inside a tube: results}

We now  have the tools necessary to characterise the locomotion of squirmers inside a tube. Our computational results, presented in this section, are organised as follows. We first compute the swimming kinematics and power consumption of a squirmer instantaneously located at various positions inside the tube while its orientation is kept parallel to the tube axis. These results then enable us to understand the origin of the two-dimensional wave-like trajectory for a neutral squirmer inside the tube.  We also  analyse the asymptotic stability of trajectories close to solid walls  \citep{yizhar_pre_stability,Darren_pre_point}. We then move on to examine the general three-dimensional helical trajectory of a neutral squirmer and also consider the kinematics of pusher and puller swimmers. Finally, we study locomotion induced by normal surface deformation  and consider locomotion inside a curved tube.

\subsection{Static kinematics and energetics}\label{static}

\begin{figure}
   \centering
   \includegraphics[width=0.49 \textwidth]{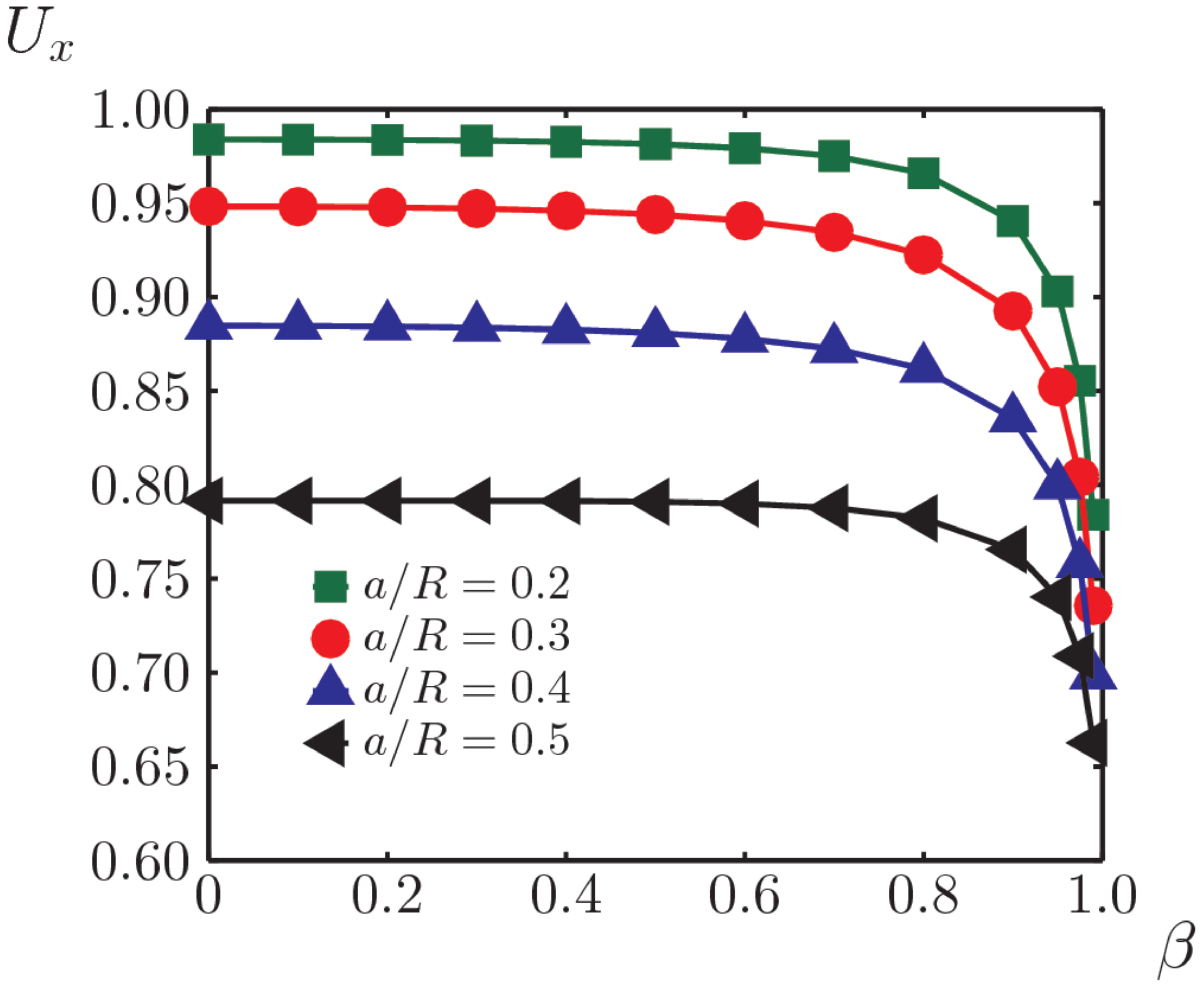} 
      \includegraphics[width=0.49
\textwidth]{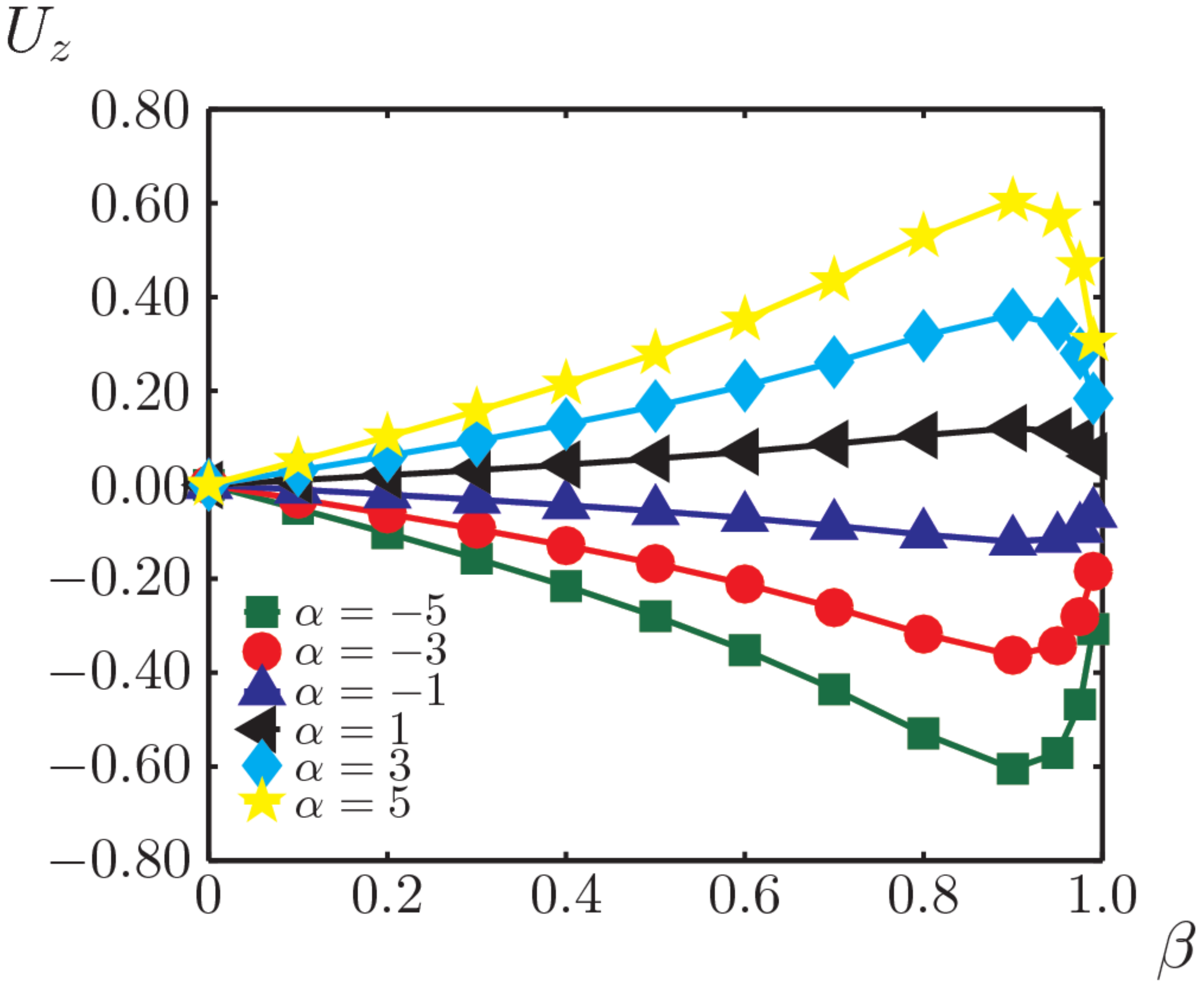}
   \caption{
Instantaneous  swimming speed for a squirmer in a capillary tube. 
Left: Swimming velocity in the axial direction, $U_{x}$(independent of $\alpha$), scaled by the swimming speed in free space $U_{ST}^{F}$; different values of $a/R$ are reported with a maximum value of $\beta=0.99$.  Right: Swimming velocity $U_{z}$ in the transverse direction, scaled by the swimming speed in free space; here $a/R$ is fixed to $0.3$. Different values of $\alpha$ are reported with maximum value of $\beta=0.99$. In both figures, the squirmer is located at $\left(0,0,-\beta\left(R-a\right)\right)$ with its orientation parallel to the tube axis.
\label{fig:spd_tan}
}
\end{figure}

To start our investigation, we first numerically calculate the swimming speed and power consumption for a squirmer exploiting pure tangential surface deformation (for
completeness, results on squirmers with normal surface deformations are shown in Sec.~\ref{jet}). 
We fix $B_{1}=1$ and vary the value of  $\alpha$, while different values of $a/R$ and $\beta$ are chosen to address the effect of confinement and eccentricity on the instantaneous swimming kinematics.

In figure~\ref{fig:spd_tan}, we plot the  instantaneous  swimming speed of a squirmer with orientation parallel to the tube axis (positive $\mathbf{x}$ direction) and location $\left(x,y,z\right)= \left(0,0,-\beta\left(R-a\right)\right)$. The swimming velocity parallel to the tube axis ($U_{x}$) is displayed in  figure~\ref{fig:spd_tan} (left) while the velocity perpendicular to it ($U_{z}$) is shown in figure~\ref{fig:spd_tan} (right). {Interestingly, both pushers and pullers have the same swimming speed, $U_{x}$, as  the neutral squirmer. This is due to the fact that the second squirming mode, $\sim B_{2}\sin2\theta$, is front back symmetric, and thus produces zero wall-induced velocity \citep{ericPRL_08}, as confirmed by our simulation.} We observe  numerically that when $\alpha=0$, there is only one non-zero velocity component, namely $U_{x}$. In contrast, for pushers and pullers ($\alpha \neq0$) a non-zero transverse velocity component, $U_{z}$, is induced. The value of $U_{x}$ is seen to decrease with confinement, $a/R$, and eccentricity, $\beta$, as shown in figure~\ref{fig:spd_tan} (left). The sharp decrease when  $\beta$ is beyond $\approx 0.8$ is due to the strong drag force experienced closer to the wall which overcomes the propulsive advantage from near-wall locomotion.

The transverse velocity, $U_{z}$,  shown in figure~\ref{fig:spd_tan} (right), is plotted against the swimmer position,  $\beta$, for different values of $\alpha$ while the confinement is fixed at $a/R=0.3$.  In the case of a puller ($\alpha>0$), the swimmer will move away from the nearest wall ($U_{z}>0$) while a
pusher ($\alpha<0$) will move towards the
nearest boundary ($U_{z}<0$), as expected  considering the dipolar velocity field generated by  
squirmers (see also Sec.~\ref{puller_traj}). The absolute value of of 
$U_{z}$ increases with $\alpha$ and is of the same magnitude for  pushers and  pullers of equal and opposite strength. A similar effect  was explained in~\citet[]{ericPRL_08} for a plane wall, although in that case, the cell was approximated by a point stresslet and the cell-wall distance was considerably larger than the cell size.  By probing hydrodynamics very close to the wall, we observe that  the magnitude of $U_{z}$ does actually not vary monotonically with $\beta$, instead reaching a maximum value as $\beta\approx
0.9$. Moving away from the tube centre, the transverse velocity increases due to stronger hydrodynamic interactions with the tube walls before decreasing owing to a significantly larger hydrodynamic resistance very close to the tube boundaries.

\begin{figure}
   \centering
   \includegraphics[width=0.49 \textwidth]{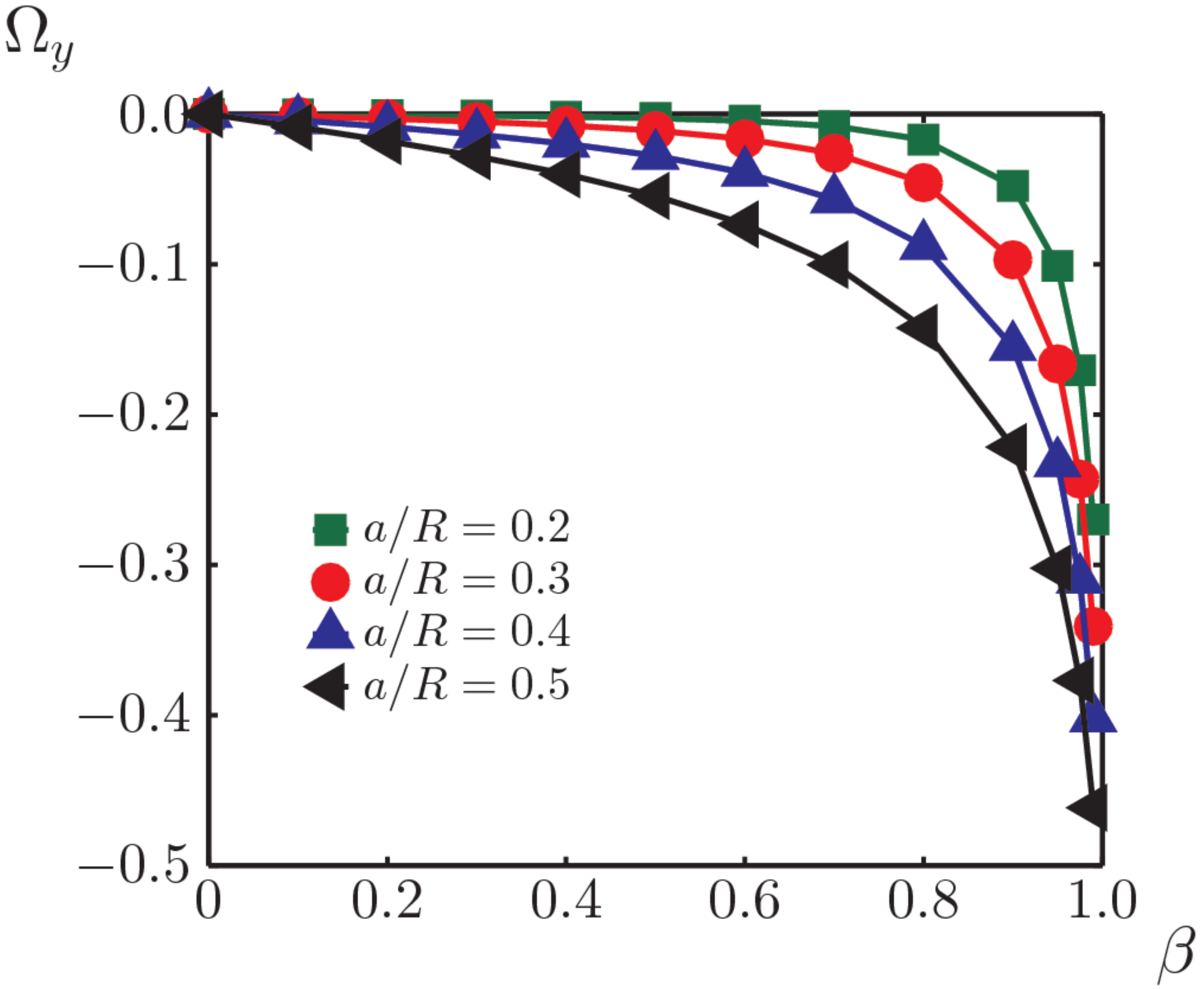} 
   \includegraphics[width=0.49 \textwidth]{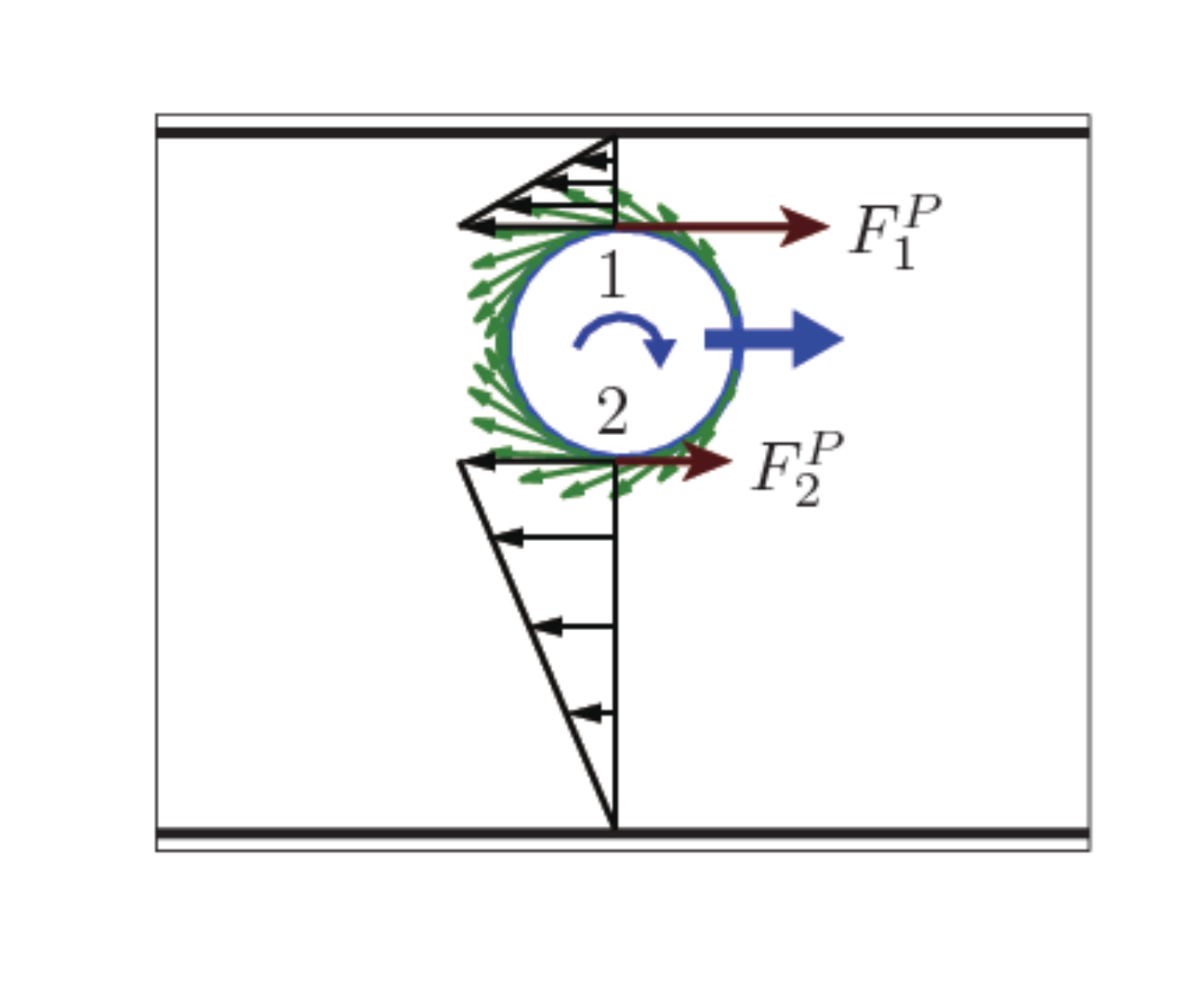} 
   \caption{Left: Rotational velocity  of the squirmer in the direction normal to the plane of
locomotion, $\Omega_{y}$. The squirmer is located at $\left(0,0,-\beta\left(R-a\right)\right)$ with its orientation parallel to the tube axis. Different values of $a/R$ are plotted with maximum value of $\beta=0.99$.
Right: Physical picture of cell rotation near the walls. The circle indicates the spherical squirmer and the green arrows denote the tangential
surface velocity imposed by the squirmer. Blue straight arrow and curved arrow denote the cell
orientation and rotational velocity respectively; $1$ indicates the closest point on the cell to the top
wall and $2$ the closest point to the bottom wall while
$F^{P}_{1}$ and $F^{P}_{2}$ are the wall-induced shear  forces generated near point $1$ and $2$.}
   \label{fig:omegay_tan}
      \label{fig:squirmer_turn_explain}
\end{figure}

Beyond the translational velocities, the squirmers also rotate due to hydrodynamic interactions with the tube boundaries. Numerical results show that the magnitude of the rotational velocity, $\boldsymbol{\Omega}$, is independent on the  dipole strength, $\alpha$, and that all squirmers rotate away from the closest wall. {This is also attributed to the front-back symmetric distribution of the second squirming mode.} Using our notation, we therefore obtain that  squirmers rotate in the $-y$
direction. The value of  $\Omega_{y}$ is displayed in figure~\ref{fig:omegay_tan} (left). Its magnitude increases with eccentricity, $\beta$,
and confinement, $a/R$, as a result of stronger hydrodynamic interactions. To
explain the sign of the rotational velocity, we look in detail at a neutral squirmer in figure~\ref{fig:squirmer_turn_explain} (right), in the case where the swimmer is located closer to the top wall. Green arrows display the tangential surface deformation which generates locomotion. Given points $1$ and $2$ on the squirmer surface, the black arrows
indicate the velocity field and show that the shear rate is higher near point $1$ than point $2$.
Consequently, the wall-induced force on point $1$, $F^{P}_{1}$,  is larger than that on point $2$,
$F^{P}_{2}$, producing a resultant clockwise torque. Since the total torque on the squirmer
is zero, the squirmer has to rotate in the clockwise direction to balance this torque, escaping from the
top wall. When the squirmer is closer to the top wall, an increased asymmetry will induce a
stronger rotation.

\begin{figure}
\centering
\subfigure[ $\quad \alpha=0$]{
   \includegraphics[scale = 0.17] {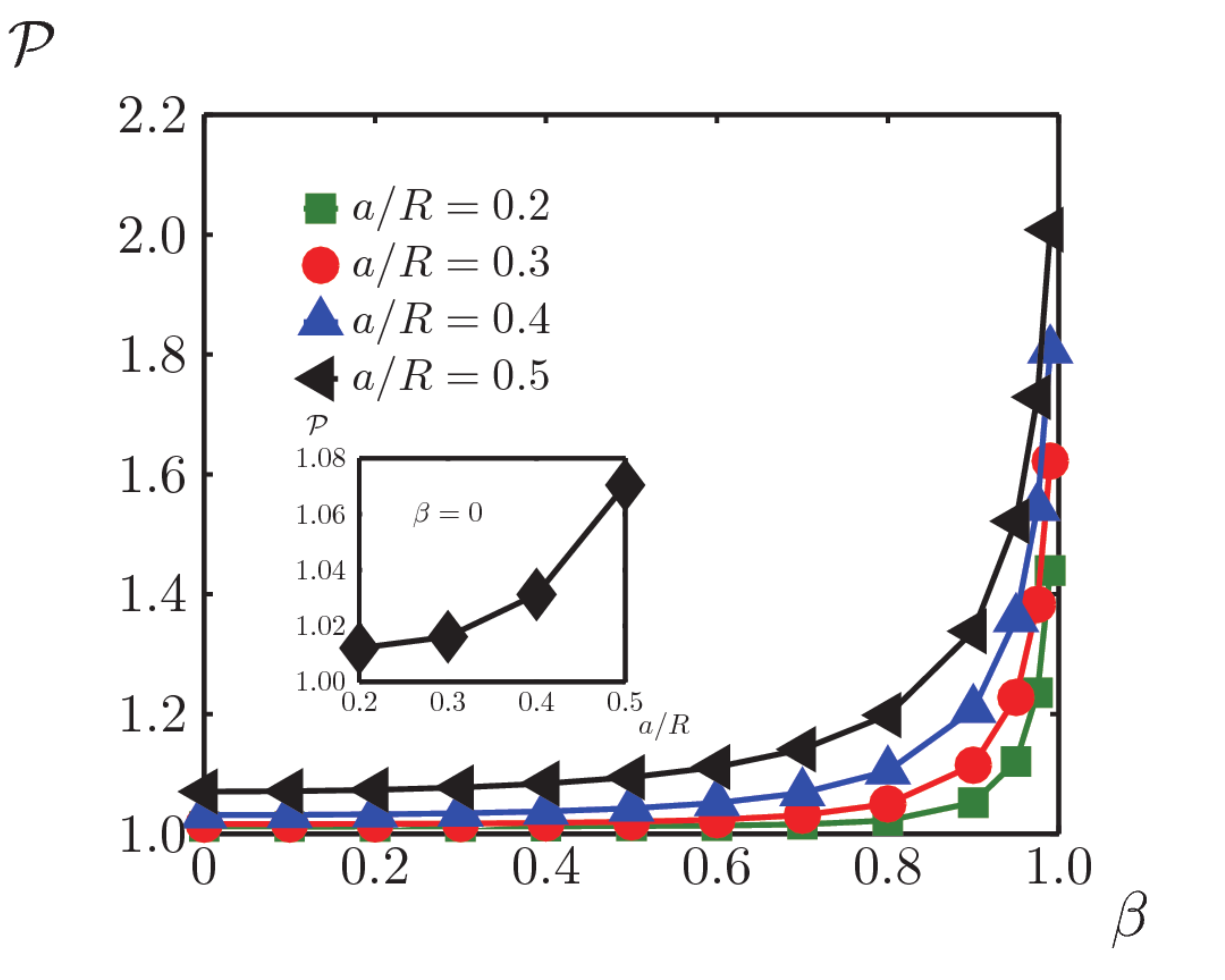}
   \label{subfig:pow_tanB2_0}
 }
 \subfigure[ $\quad \alpha=5$]{
   \includegraphics[scale = 0.17] {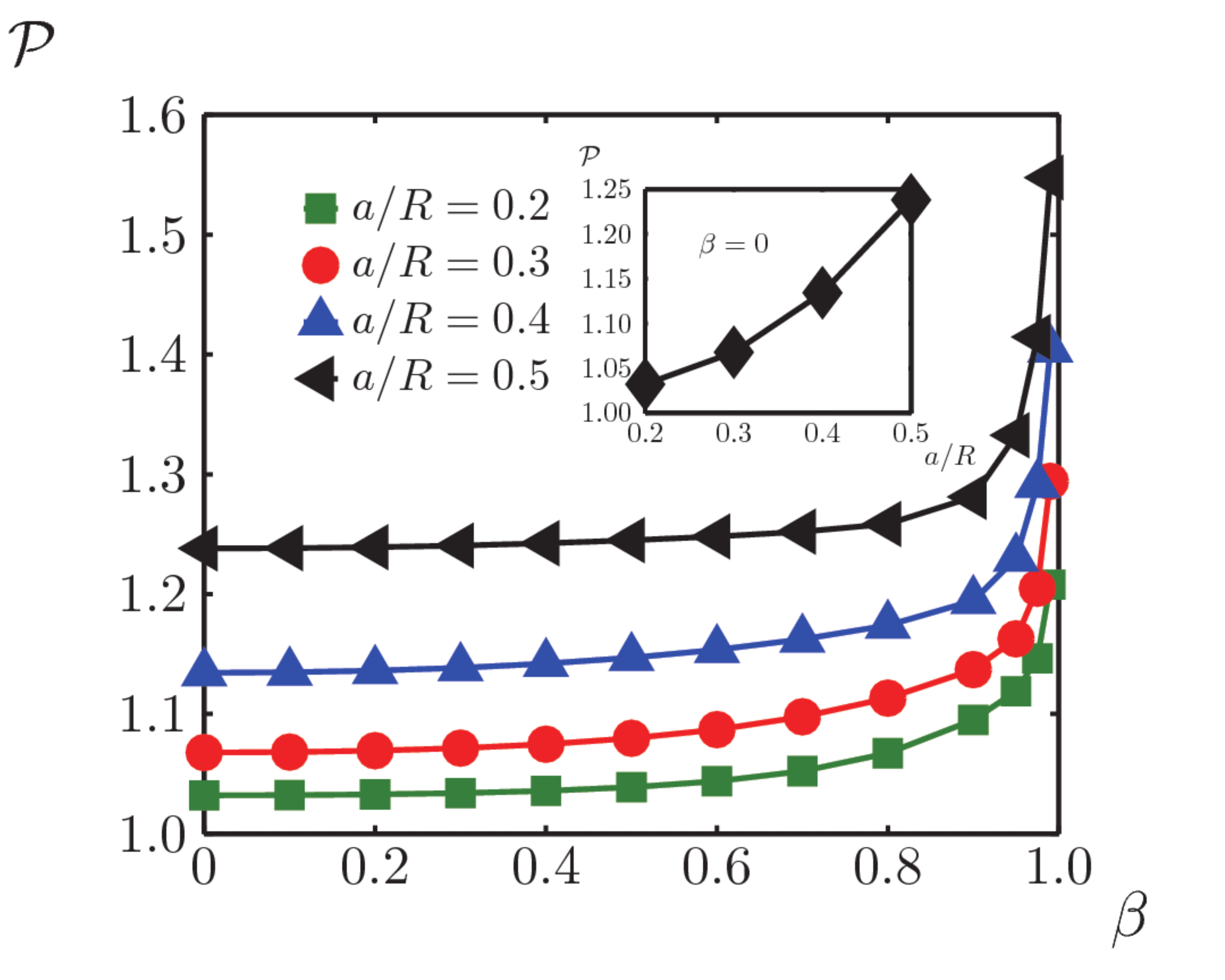}
   \label{subfig:pow_tanB2_5}
 }
\caption{Power consumption, $\mathcal{P}$, of the neutral squirmer ($\alpha=0$, left) and puller ($\alpha=5$, right), scaled by their corresponding values in free space. Insets display $\mathcal{P}$ as a function of $a/R$ in the case $\beta=0$. The orientation of the squirmer is parallel to the tube axis with the maximum value of $\beta=0.99$.}
\label{fig:pow_tan}
\end{figure}

Next, we analyse the power consumption by the squirmer. The power, $\mathcal{P}$, is defined as $\mathcal{P}=\int_{S}\mathbf{f}_{out} \cdot \mathbf{u}_{S}dS$, where $\mathbf{f}_{out}$ is the force per unit area exerted from the outer surface of the body onto the fluid and $\mathbf{u}_{S}$ is the squirming velocity. In the single-layer potential formulation, equation~\ref{equ:stokeslet} as in~\citet[]{swimInter-Pedley}, the unknown $\mathbf{f}$ is the sum of the force density from outer ($\mathbf{f}_{out}$) and inner ($\mathbf{f}_{in}$) surface. We therefore rewrite the power as $\mathcal{P}=\int_{S}\mathbf{f} \cdot \mathbf{u}_{S}dS - \int_{S}\mathbf{f}_{in} \cdot \mathbf{u}_{S}dS$, where $\int_{S}\mathbf{f}_{in} \cdot \mathbf{u}_{S}dS$ denotes the viscous dissipation of the flow inside the squirmer. We thus need to subtract the internal viscous dissipation in the fluid given by the numerics where $\mathbf{f}_{in}$ can be derived analytically based on the squirming velocity. In figure~\ref{fig:pow_tan}, we depict the dependence  of $\mathcal{P}$, scaled by the corresponding value in free space, with $\beta$ and for different values of $\alpha$.  For each gait, $\mathcal{P}$ increases slowly until $\beta\approx0.8$ followed by a rapid increase for cells closer to the wall. Such a drastic power increase is in agreement with the sharp decrease in swimming speed close to the tube,  and {consequently, a significant decrease in swimming efficiency is expected.} In addition, as the confinement is getting stronger, the eccentricity of the swimmer's position becomes more important. For example, as $\beta$ changes from $0$ to $0.99$, the power consumption of a neutral squirmer $\mathcal{P}$ increases only by around $45\%$ for $a/R=0.2$ but by $85\%$ for $a/R=0.5$.

\subsection{Two-dimensional wavelike motion of the neutral squirmer}\label{2dwaveneu}

\begin{figure}
   \centering
   \includegraphics[width=0.9 \textwidth]{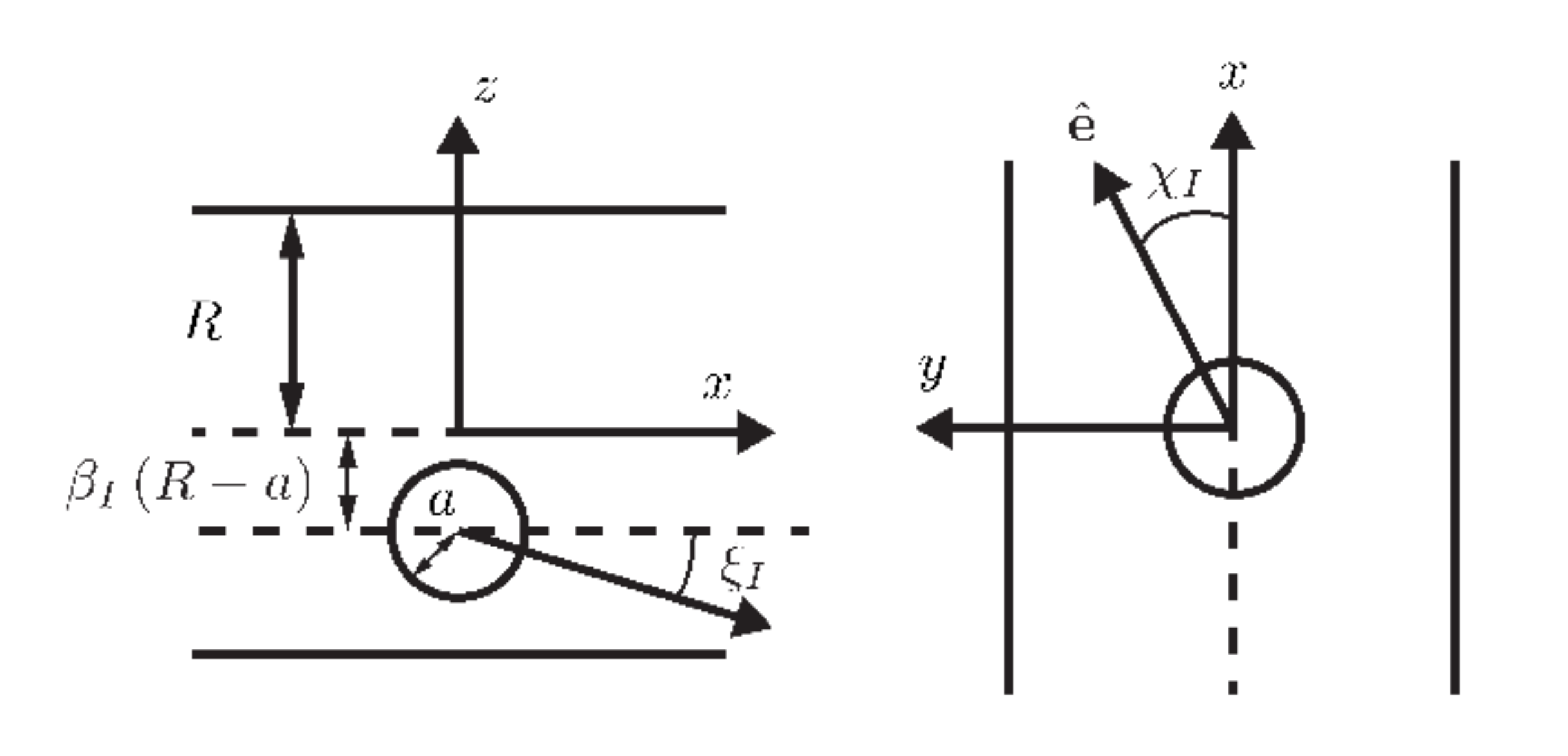}
   \caption{Sketch of the spherical squirmer inside the tube with coordinate system and angles.
The initial dimensionless off-axis distance is measured by $\beta_{I}$ while the angles $\xi_{I}$ and $\chi_{I}$ control the initial cell orientation, $\hat{\mathbf{e}}$. For $\chi_{I}=0$, and in the absence of noise,  the squirmer motion is restricted to the $x-z$ plane.}
   \label{fig:orient}
\end{figure}

We next study in detail the trajectory of a squirmer inside a tube with fixed confinement; unless otherwise stated, all  results in this section use the same value, $a/R=0.3$.  {The cell is neither a pusher nor a puller, but a neutral squirmer generating potential flow field ($\alpha=0$).} The initial position and orientation of the cell are defined as in figure~\ref{fig:orient}. The cell is initially placed at $(0,0,-b_{I})$, with $b_{I}=\beta_{I}\left(R-a\right)$, and oriented parallel to the axis ($\xi_{I} = 0$); the motion of the cell will also be restricted to the $x-z$ plane ($\chi_{I}=0$). We calculate the translational and rotational velocity of the cell at each time step and update its position using fourth-order Adams-Bashforth scheme as in  \citet[]{IshikawaUnsteady}. Note that in the simulations the cell always remains in the centre of the computational domain (while its axial velocity is stored for post-processing), which allows to minimise the error introduced by domain truncation.

\begin{figure}
   \centering
   \includegraphics[width=0.95 \textwidth]{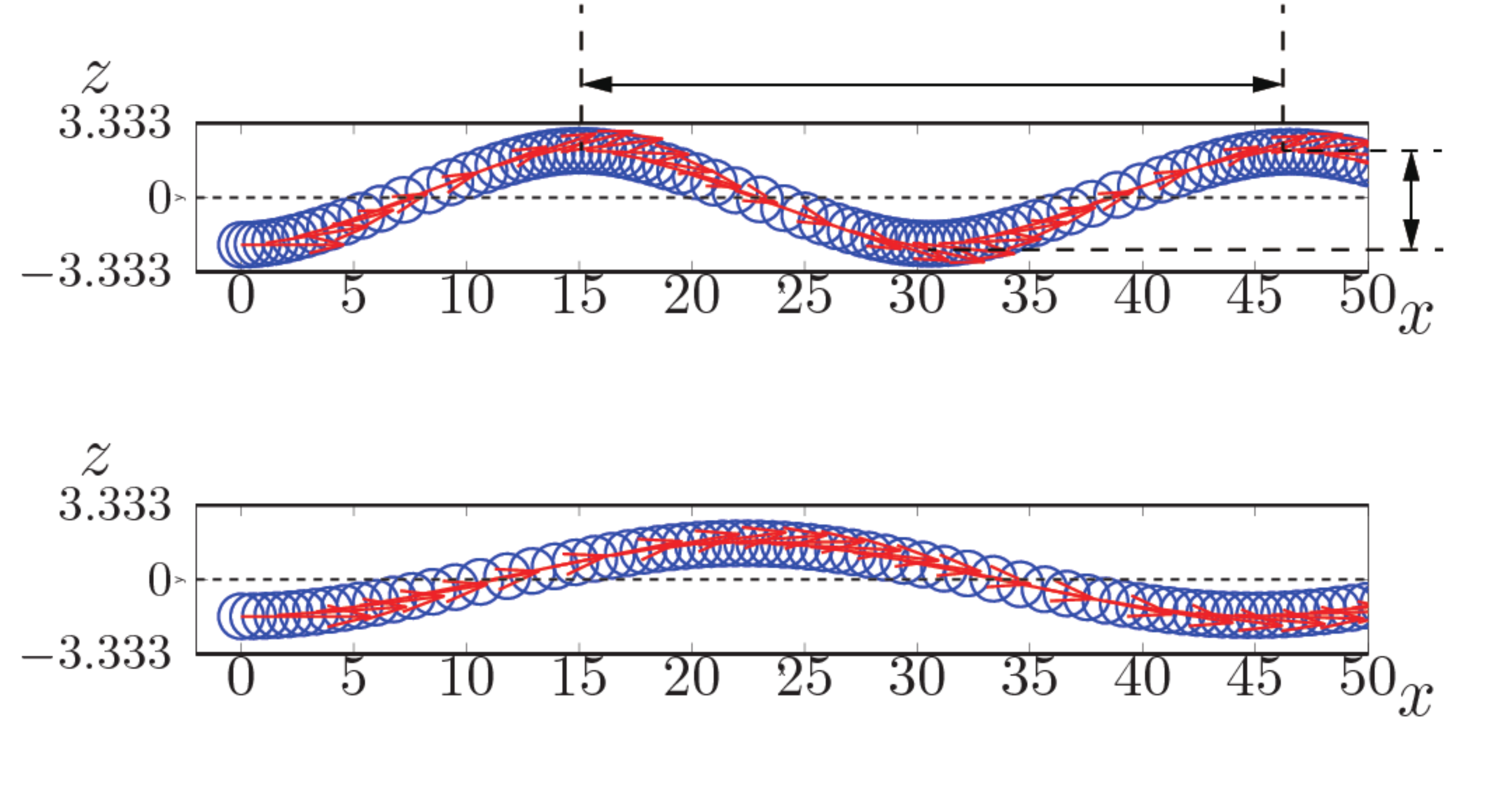}
   \put(-13,140){\Large{$A$}}
   \put(-145,180){\Large{$\lambda$}}
   \put(-320,90){\Large{$\beta_{I}=0.7$}}
   \put(-320,180){\Large{$\beta_{I}=0.9$}}
   \caption{Two-dimensional trajectories of a neutral squirmer inside a capillary tube with confinement $a/R$=0.3. All positions are measured in  units of cell radius $a$ (same for figures hereinafter unless otherwise specified). Blue circles and red arrows indicate, respectively,  the instantaneous position and orientation of the squirmer. The cell is released from $(0,0,-\beta_{I}\left(R-a\right))$ with $\beta_{I}=0.9$ (up) and $\beta_{I}=0.7$ (bottom), while the initial orientation is parallel to the axis. We denote $\lambda$  the wavelength of the periodic trajectory and $A$ its amplitude.}
   \label{fig:traj_show}
\end{figure}
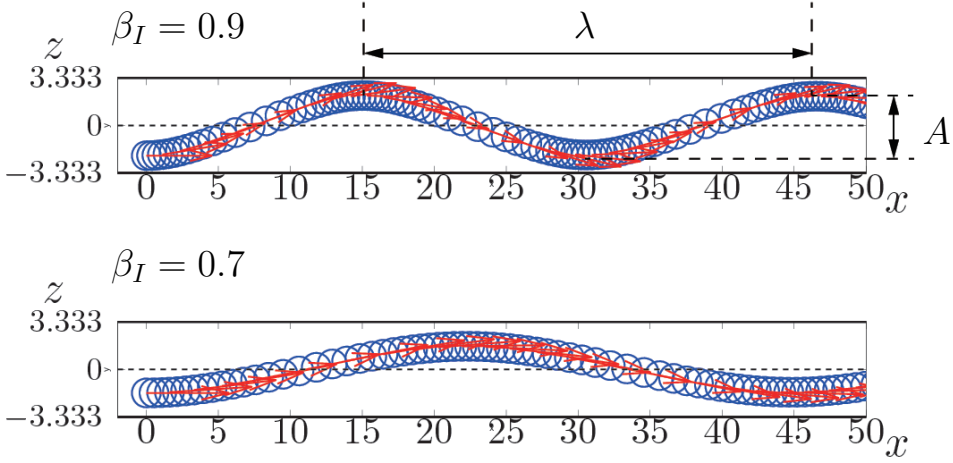

\begin{figure}
   \centering
   \includegraphics[width=0.6\textwidth]{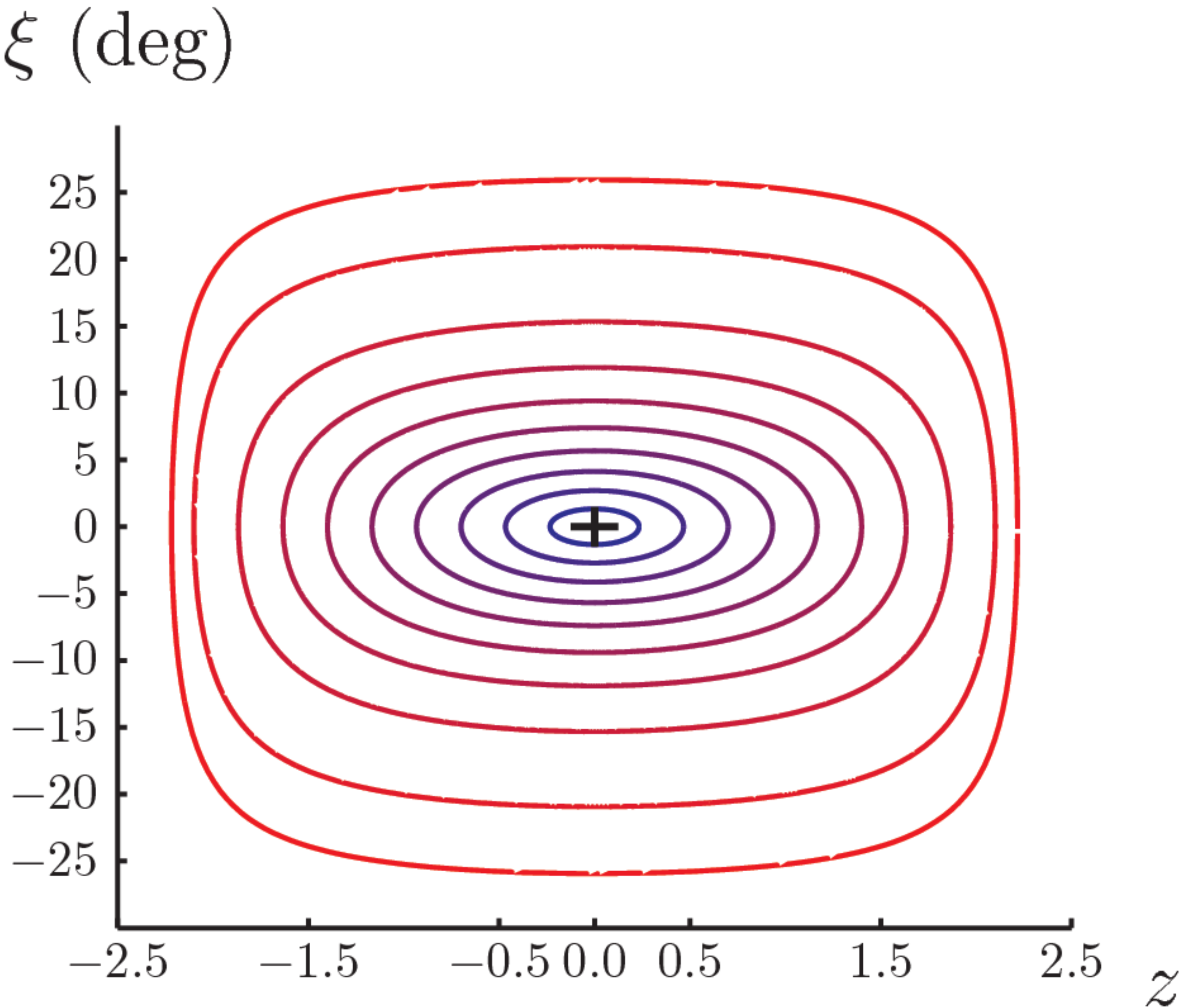}
   \caption{Phase portrait for the neutral squirmer inside the tube in the $\left(z,\xi \right)$ plane
with confinement $a/R=0.3$. Closed orbits correspond to $2$D wavelike trajectories. The black
cross denotes the equilibrium point $\left(z,\xi\right)=\left(0,0\right)$.}
   \label{fig:phase_atoR0.3}
\end{figure}

Our computations show that the squirmer always displays a periodic wavelike trajectory in the tube, with amplitude $A$ and wavelength $\lambda$. This is illustrated in figure~\ref{fig:traj_show} for $\beta_I=0.9$ (top) and $\beta_I=0.7$ (bottom). The wave amplitude does not change over time and is two times the initial off-axis distance, namely, $A=2b_{I}$. The presence of a nonzero rotational velocity, ${\Omega}_y$, discussed above and shown in figure~\ref{fig:omegay_tan}, is the key parameter leading to the periodic trajectory. By considering cases where the initial orientation of the cell is not parallel to the axis (thus for which the orientation vector has non-zero $x$ and $z$ components) and we find that as long as the squirmer does not immediately descend into the wall, a wavelike trajectory is also obtained. To present all results in a concise manner, we consider the motion of the neutral squirmer as a dynamical system similarly to recent work on two-dimensional swimming~\citep{yizhar_pre_stability, darren_gap}. The trajectory is defined by two parameters,  the off-axis distance ($z$) and  the angle between the swimmer orientation and the tube axis ($\xi$). We report the phase portrait of the neutral squirmer in the $\left(z,\xi\right)$ plane in figure~\ref{fig:phase_atoR0.3}, where the solid curves show the trajectories. The marginally stable  point $\left(0,0\right)$ corresponds to locomotion along the axis of the tube. For any initial conditions $\left(z,\xi\right)$, the neutral squirmer swims along wavelike trajectories corresponding to the periodic orbits in figure~\ref{fig:phase_atoR0.3} (the largest periodic orbit in the figure has a maximum $\beta$ of $0.95$).

\begin{figure}
   \centering
   \includegraphics[width=0.8 \textwidth]{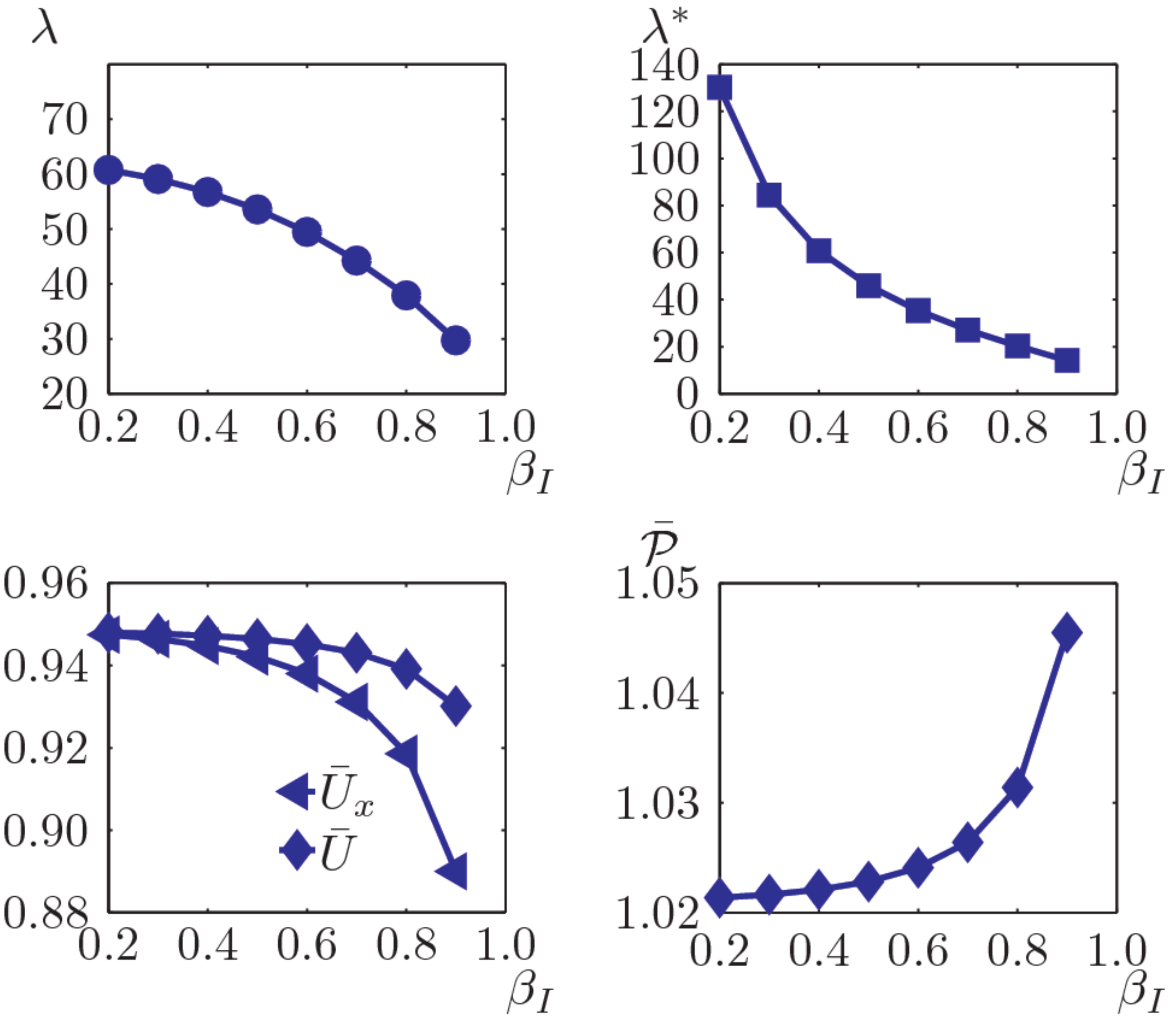}
   \caption{Dynamics and kinematics of the neutral squirmer in a tube as a function of the initial dimensionless off-axis position,  $\beta_{I}$. Top left: wavelength, $\lambda$, of the periodic trajectory. Top right: wavelength-to-amplitude ratio,  $\lambda^{*}=\lambda/(A/2)$. Bottom left: time-averaged swimming speed in the axial direction, $\bar{U}_{x}$,  and along the trajectory, $\bar{U}$, both rescaled by the  free-space swimming velocity. Bottom right: time-averaged power consumption $\bar{\mathcal{P}}$, rescaled by  value in free space.}
   \label{fig:waveinfo_aToR0.3}
\end{figure}

The  main characteristics of the squirmers' trajectories are shown in figure~\ref{fig:waveinfo_aToR0.3} for different initial positions, $\beta_{I}$. We display the trajectory wavelength, $\lambda$, and the wavelength-to-amplitude ratio,  
 $\lambda^{\ast}={\lambda}/({A/2})$. It is clear that $\lambda$ and $\lambda^{\ast}$ both decrease with $\beta_{I}$. Indeed, when the swimmer is at the crest or trough of the periodic trajectory, stronger rotation occurs for larger $\beta_{I}$. Therefore, the swimmer will escape from the nearest wall more rapidly, resulting in a decrease of the wavelength. We also show in figure~\ref{fig:waveinfo_aToR0.3} that the time-averaged axial speed, $\bar{U}_{x}$, and
the time-averaged swimming speed along the trajectory, $\bar{U}$, decrease with $\beta_{I}$ whereas the time-averaged power consumption, $\bar{\mathcal{P}}$, increases when the squirmers move closer to the wall. 

\subsection{Three-dimensional helical trajectory of the neutral squirmer}
\label{3dspiralneu}

\begin{figure}
   \centering  
      \includegraphics[width=0.9 \textwidth]{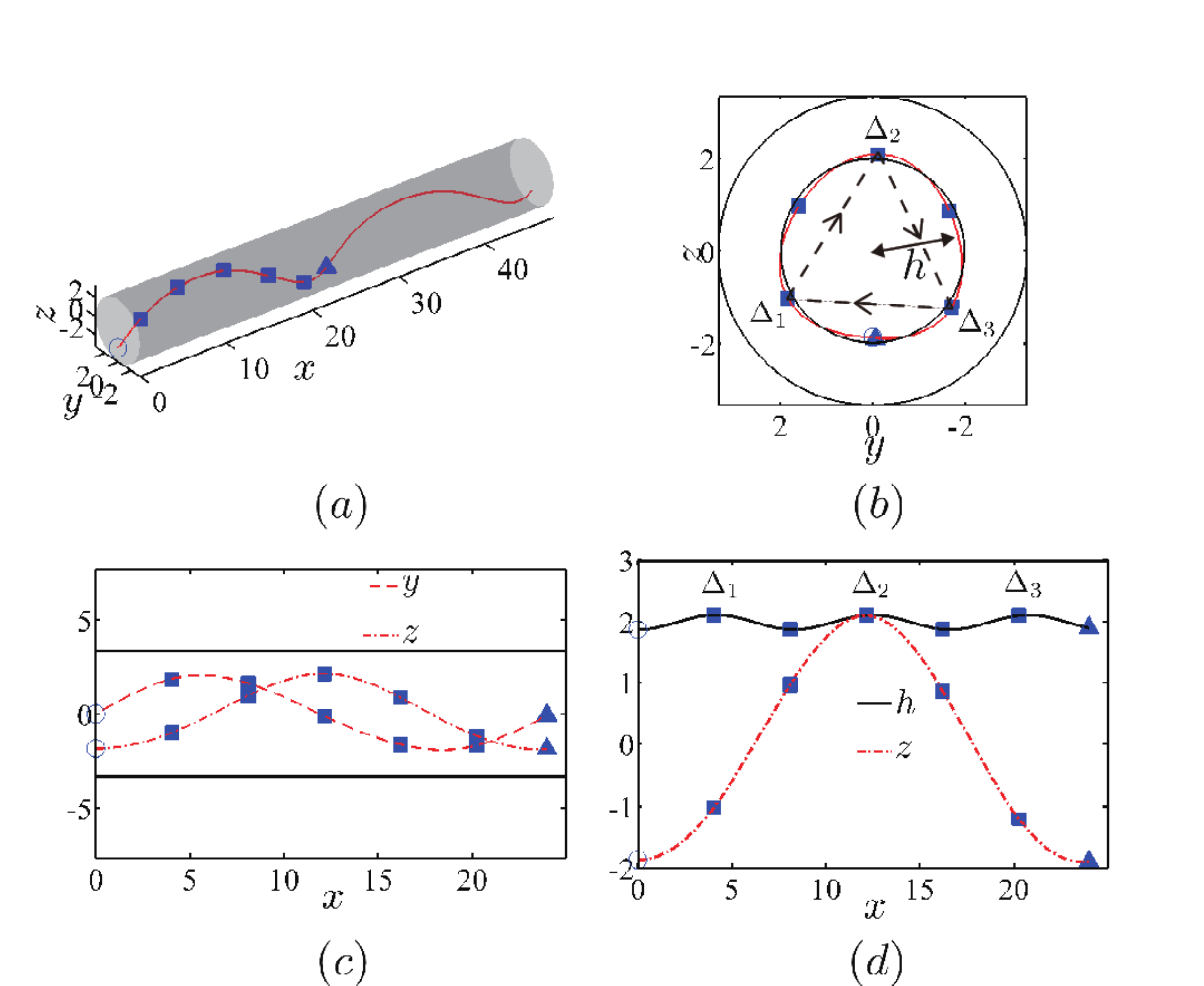}
\caption{Three-dimensional trajectory of the neutral squirmer in the tube. The initial
position and orientation are given by $\beta_{I}=0.8$, $\chi_{I}=30^{\circ}$, $\xi_{I}=0^{\circ}$. (a): 
trajectory in perspective view; the empty circle and solid triangle indicate the start and end of one periodic orbit; 
(b): trajectory in the $y-z$ plane (axes shown in figure~\ref{fig:neu_helix}a);  
(c): trajectories in  the $x-y$ (dashed) and $x-z$ (dot dashed) planes;  (d): relation between the wavelike motion developed in the axial and azimuthal direction.}
   \label{fig:neu_helix}
\end{figure}

By tilting the initial cell orientation, $\hat{\mathbf{e}}$,  off the $x-z$ plane, the squirmer trajectories become three dimensional and take the shape of a helix, a feature we address in this section. As in the two-dimensional case, these three-dimensional trajectories are a consequence of hydrodynamics interactions only. Recent experiments in \citet[]{sunny_para_tube} showed that \textit{Paramecium} cells display helical trajectories when swimming inside capillary tubes, a feature our simulations are thus able to reproduce. Note that some \textit{Paramecium} cells also follow helical trajectories  in free space due to asymmetries in the shape of their body and the beating of its cilia. In the current work we focus on swimmers whose helical dynamics arises only in  confinement.

\begin{figure}
\centering
\includegraphics[width=0.9\textwidth]{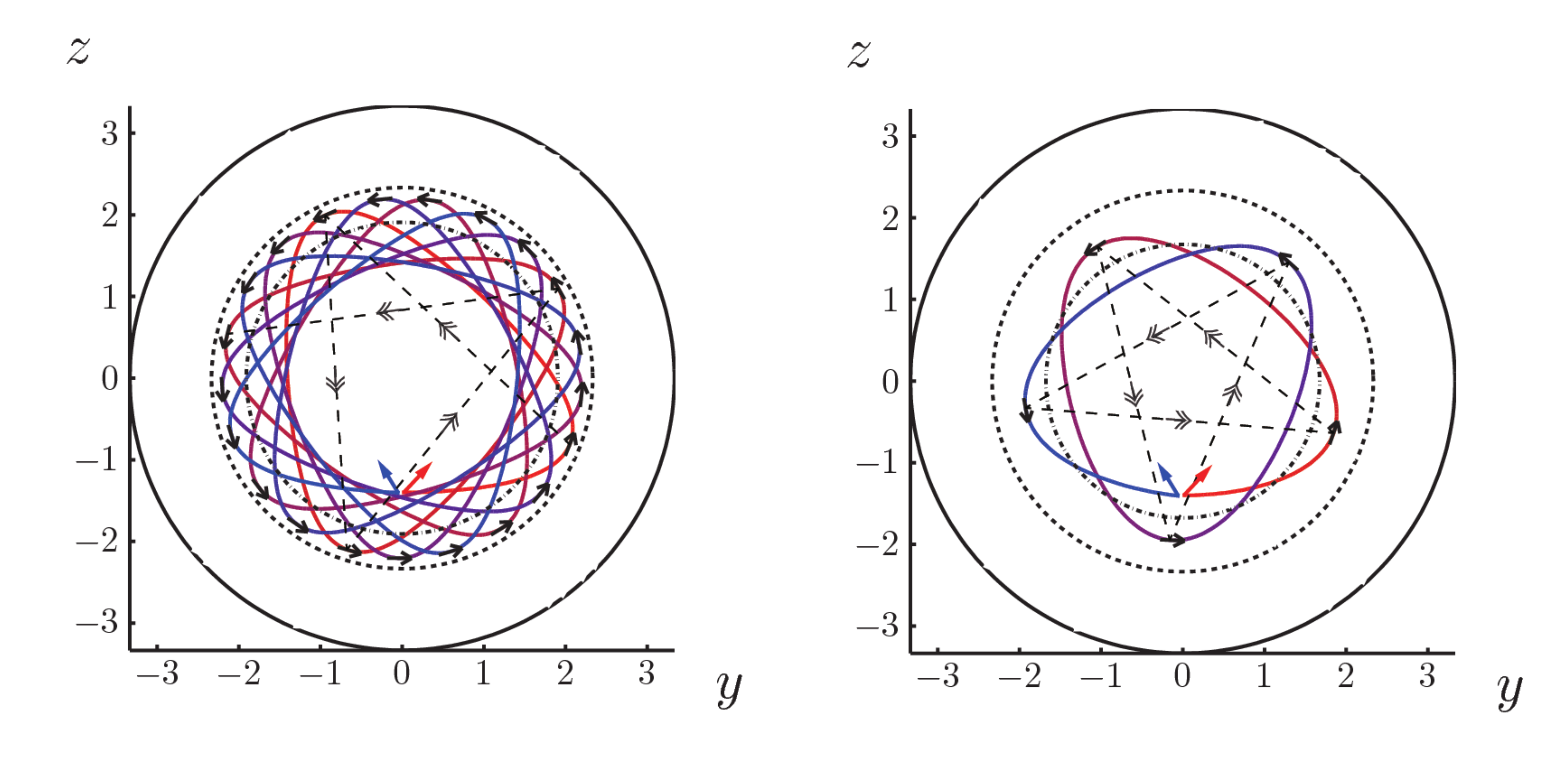}
   \put(-273,72){\textcolor{blue}{\tiny{$\mathrm{End}$}}}
   \put(-257,72){\textcolor{red}{\tiny{$\mathrm{Start}$}}}
   \put(-99,72){\textcolor{blue}{\tiny{$\mathrm{End}$}}}
   \put(-81,72){\textcolor{red}{\tiny{$\mathrm{Start}$}}}
   \put(-285,5){\footnotesize{$\beta_{I}=0.6\;\chi_{I}=30^{\circ}$}}
   \put(-112,5){\footnotesize{$\beta_{I}=0.6\;\chi_{I}=20^{\circ}$}}
 \caption{Periodic orbits of the neutral squirmer in  the transverse plane for two different initial positions and orientations. Left: $\left(\beta_{I}, \chi_{I}\right) =\left(0.6, 30^\circ\right)$. Right: $\left(\beta_{I}, \chi_{I}\right) =\left(0.6, 20^\circ\right)$.}
   \label{fig:helix_beta0.6angle30}
\end{figure}

We introduce $\chi_{I}$ as the yaw
angle between the initial cell orientation and the $x-z$ plane (see figure~\ref{fig:orient}), so that the initial orientation becomes $\left(\cos\left(\chi_{I}\right), \sin\left(\chi_{I}\right), 0\right)$. In our simulations, $\beta_{I}$ ranges from
$0.3$ to $0.9$ and $\chi_{I}$  from $20^\circ$ to $40^\circ$. Within these parameters, squirmers always display helical trajectories. One such helix is plotted in figure~\ref{fig:neu_helix}, for an initial position $\beta_{I}=0.8$ and a yaw angle $\chi_{I}=40^\circ$. The helical trajectory is a combination of wavelike motions developed in the azimuthal $y-z$ plane and in the axial direction, see figure~\ref{fig:neu_helix}b and c. In figure~\ref{fig:neu_helix}b, we show the projected circular  trajectory of the swimmer in the  $y-z$ plane. In figure~\ref{fig:neu_helix}c, we show that the curves $y\left(x\right)$ and
$z\left(x\right)$ share the same wavelength and time period. We then plot the values of $z$ and $h$ (cell off-axis distance) as a function of the axial position, $x$, during one period in figure~\ref{fig:neu_helix}d to show that the wave
frequency of $h\left(x\right)$ is three times that of
$z\left(x\right)$. Indeed, the trajectory projected in the plane perpendicular to the tube axis resembles a regular triangle
 ($\Delta_{1}\Delta_{2}\Delta_{3}$), with vertices corresponding to locations of maximum off-axis distance where the cell bounces back inside the tube. In this particular case, the cell bounces off the wall three times during one orbit with an angle $\psi=60^{\circ}$. A variety of other wave patterns can be observed for different initial cell positions and yaw angels $\left(\beta_{I},\chi_{I}\right)$. We display two of them in figure~\ref{fig:helix_beta0.6angle30} in the $y-z$ plane, with $\left(\beta_{I},\chi_{I}\right)=\left(60,
30^{\circ}\right)$  (figure~\ref{fig:helix_beta0.6angle30}, left) and  $\left(60, 20^{\circ}\right)$ (figure~\ref{fig:helix_beta0.6angle30}, right). The swimmer on the left approaches the wall
$21$ times during one periodic orbit with $\psi=42.86^{\circ}$, whereas the example on the right displays a $5$ fold helix with $\psi=36^{\circ}$.

\begin{figure}
\begin{center}
   \includegraphics[width=0.4 \textwidth]{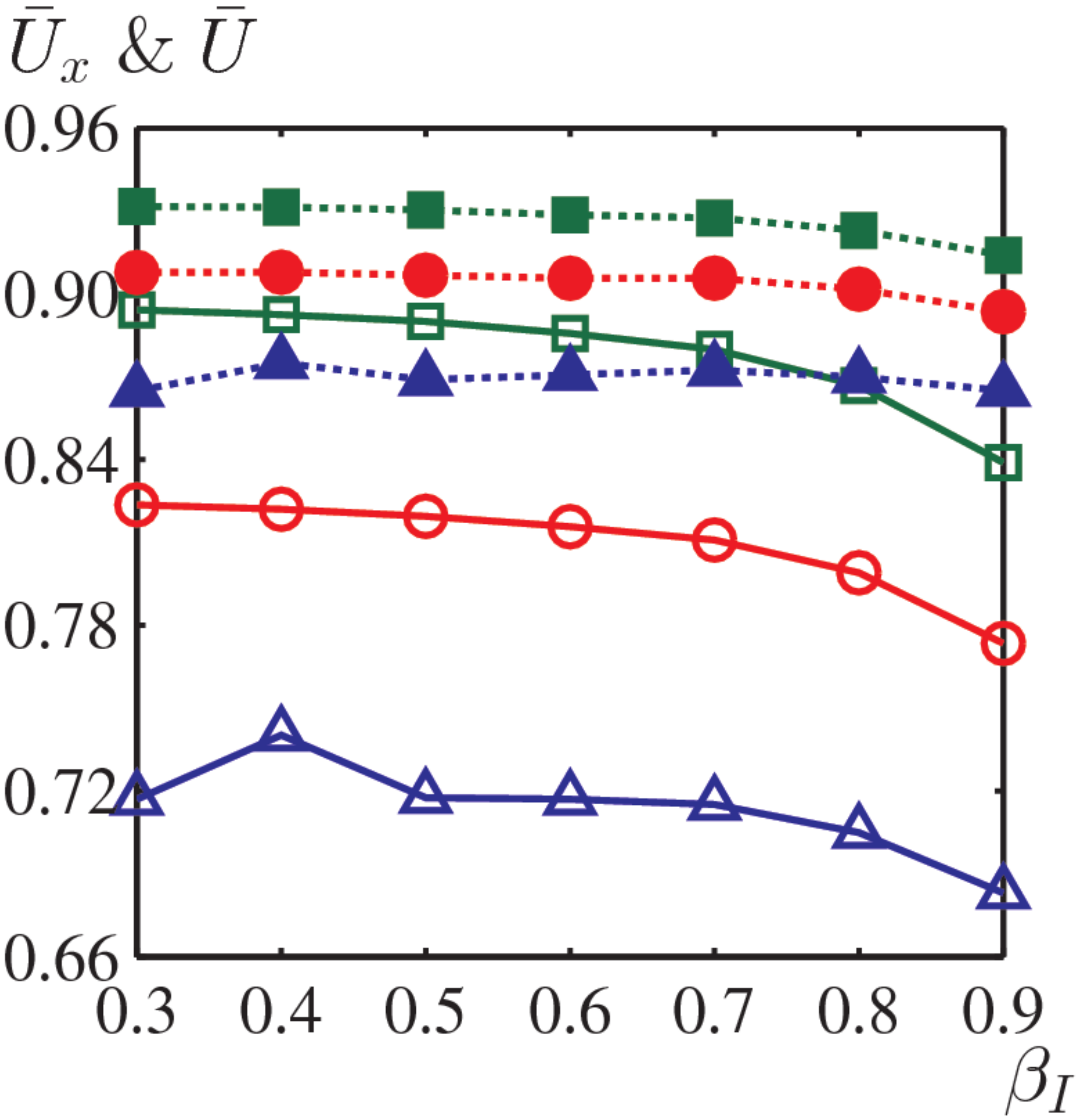}
   \put(-66,90){\textcolor{darkgreen}{\scriptsize{$\bar{U}_{x} \left(\chi_{I} = 20\right)$}}}
   \put(-55,130){\textcolor{darkgreen}{\scriptsize{$\bar{U} \left(\chi_{I} = 20\right)$}}}
   \put(-75,68){\textcolor{red}{\scriptsize{$\bar{U}_{x} \left(\chi_{I} = 30\right)$}}}
   \put(-60,110){\textcolor{red}{\scriptsize{$\bar{U} \left(\chi_{I} = 30\right)$}}}
   \put(-90,53){\textcolor{blue}{\scriptsize{$\bar{U}_{x} \left(\chi_{I} = 40\right)$}}}
   \put(-115,96){\textcolor{blue}{\scriptsize{$\bar{U} \left(\chi_{I} = 40\right)$}}}
   \includegraphics[width=0.4 \textwidth]{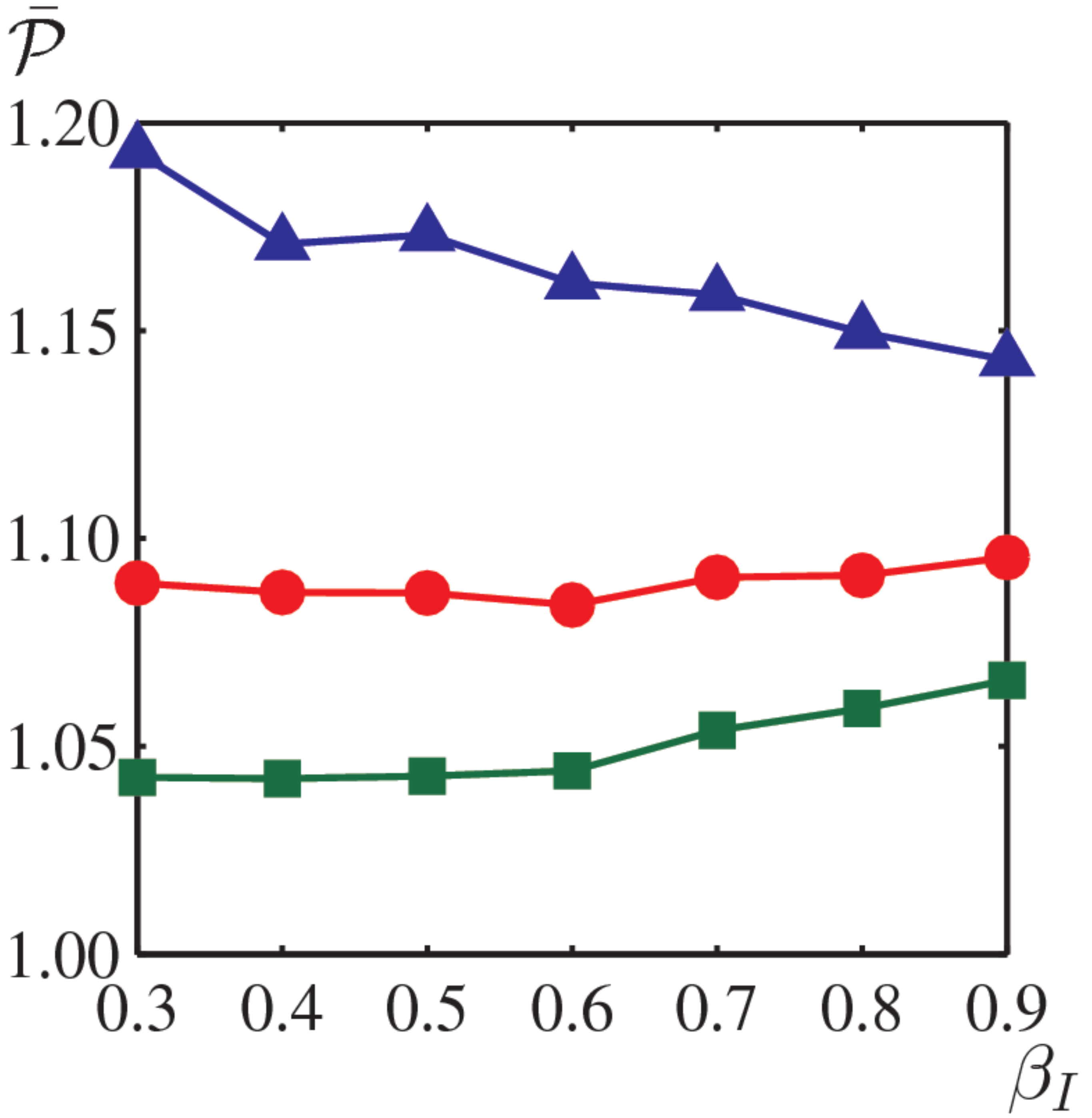}
   \put(-100,36){\textcolor{darkgreen}{\scriptsize{$\bar{\mathcal{P}} \left(\chi_{I} = 20\right)$}}}
   \put(-100,80){\textcolor{red}{\scriptsize{$\bar{\mathcal{P}} \left(\chi_{I} = 30\right)$}}}
   \put(-100,130){\textcolor{blue}{\scriptsize{$\bar{\mathcal{P}} \left(\chi_{I} = 40\right)$}}}
\end{center}
   \caption{Nondimensional time-averaged swimming speed and power consumption of squirmers with different initial position ($\beta_{I}$) and orientation ($\chi_{I}$) with three-dimensional kinematics.}
   \label{fig:spiral_u_p}
\end{figure}

Finally in figure~\ref{fig:spiral_u_p} we show the variation of the averaged swimming speed (left) and power consumption (right) with the initial cell position ($\beta_{I}$) and orientation ($\chi_{I}$), {where both the speed and power are nondimensionalized by their corresponding values in free space.} The time-averaged swimming speed along the axial direction, $\bar{U}_{x}$, and along the trajectory, $\bar{U}$, decrease clearly with $\chi_{I}$ but slowly with $\beta_{I}$. Larger values of $\chi_{I}$ and $\beta_{I}$ result in larger maximum off-axis distance, leading to higher hydrodynamic resistance from the boundaries and thus hindering locomotion. We also observe that $\bar{U}_{x}$ decreases with $\chi_{I}$ more rapidly than $\bar{U}$. As $\chi_{I}$ increases, the swimmer trajectory becomes more coiled, which significantly decreases  the swimming velocity in the axial direction. We also note that  the power consumption, $\bar{\mathcal{P}}$, increases with the the initial orientation, $\chi_{I}$, but does not change significantly with $\beta_{I}$.

\subsection{The trajectory of a puller inside the tube}\label{puller_traj}
\begin{figure}
   \centering
   \includegraphics[width=0.95
\textwidth]{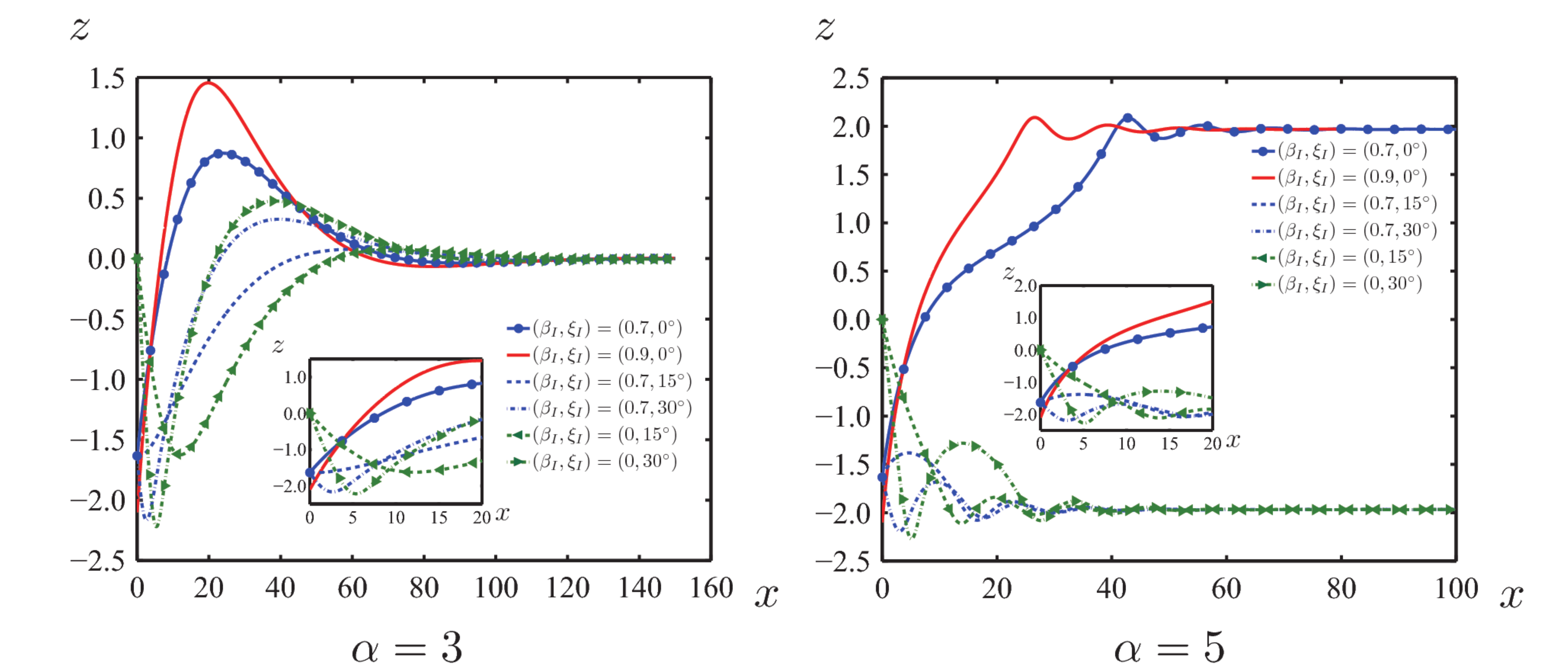}
   \caption{Two-dimensional trajectories, $z(x)$, of pullers in a tube (left: $\alpha=3$; right: $\alpha=5$). Different combinations of the initial position ($\beta_{I}$) and pitching angle ($\xi_{I}$) are chosen. The inset plots display the trajectories near the starting positions.}
   \label{fig:traj_al3_al5}
\end{figure}
\begin{figure}
   \centering
   \includegraphics[width=0.65 \textwidth]{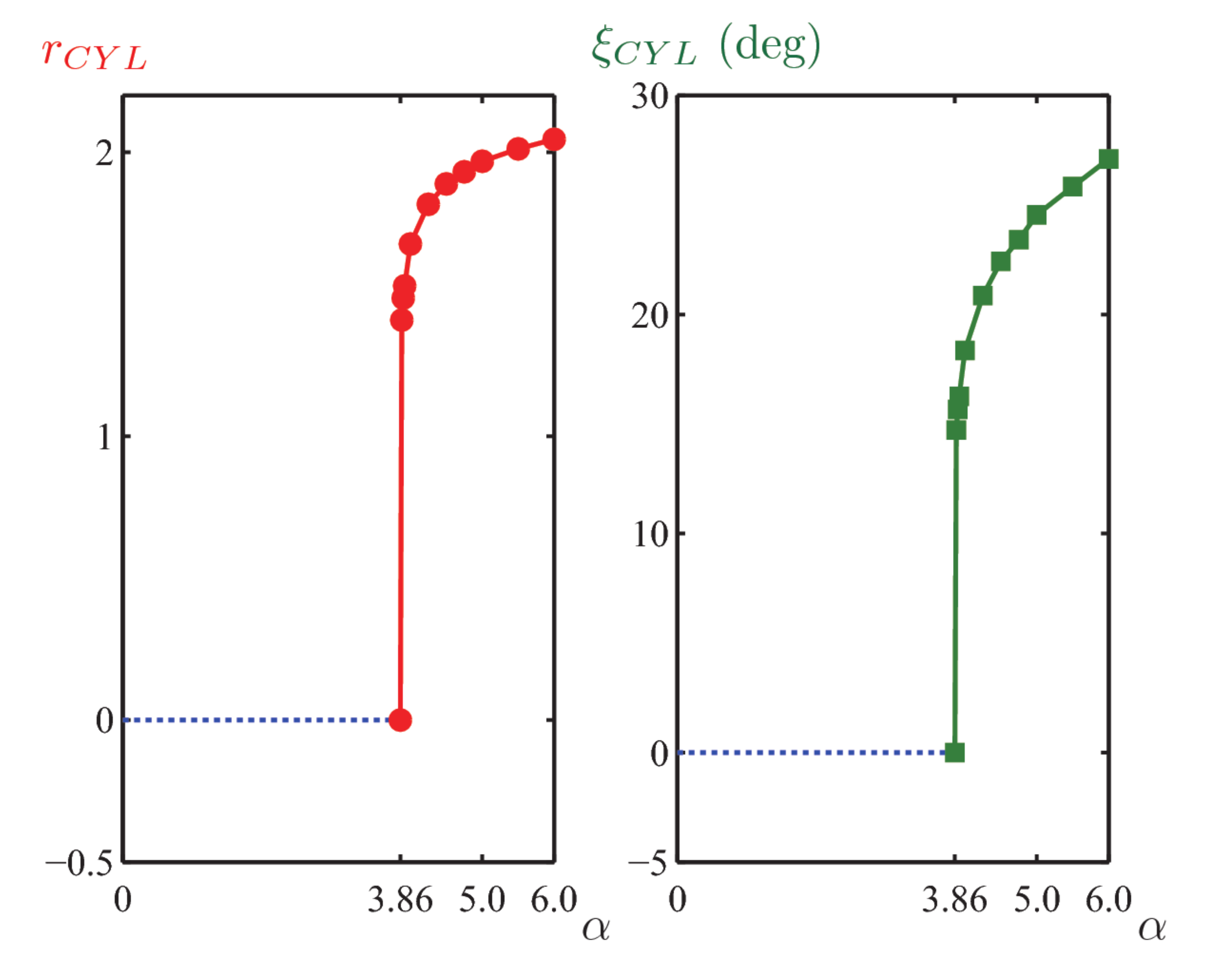}
   \caption{{Equilibrium points, $\left\{r_{CYL}, \xi_{CYL}\right\}$, of the puller in the tube as a function of the the  dipole strength, $\alpha$, and for a confinement $a/R=0.3$. Blue dashed lines ($\alpha<3.86$) show the equilibrium point at $\left(0, 0\right)$, corresponding to swimming in the  centre of the tube and along its axis. For $\alpha>3.86$, the combination $\left\{r_{CYL}, \xi_{CYL}\right\}$ characterises   the equilibrium state for swimming along a straight line with off-axis distance $r_{CYL}$ and orientation towards the wall $\xi_{CYL}$.}}
   \label{fig:pullerbranch}
\end{figure}

In this section, we study the trajectories of a puller swimmer ($\alpha>0$) in the tube. We first consider the case where the motion is restricted to the  $x-z$ plane, as in Sec.~\ref{2dwaveneu}. In figure~\ref{fig:traj_al3_al5} we show the two-dimensional trajectories of pullers having  dipole parameters of $\alpha=3$ (left) and $\alpha=5$ (right), for different initial positions, $\beta_{I}$, and orientations, $\xi_{I}$. In both cases, the swimmers initially follow wavelike trajectories with decreasing magnitude, and eventually settle  along straight trajectories, displaying thus passive asymptotic stability~\citep{yizhar_pre_stability}. {The  puller with $\alpha=3$ ends up swimming along the tube axis, with $\left(r_{CYL}, \xi_{CYL}\right) = \left(0,0\right)$ as its equilibrium point (cylindrical coordinates are used here, and $r_{CYL}$ and  $\xi_{CYL}$ denote the off-axis distance and orientation of the cell respectively).} In contrast, the puller with $\alpha=5$ swims parallel to the axis near the top or bottom wall depending on its initial position and orientation, thus its equilibrium point corresponds to swimming along an off-axis straight line. In that case, even though the trajectory is parallel to the tube axis, the swimmer remains slightly inclined towards the wall to offset the hydrodynamic repulsion from the wall. 

\begin{figure}
\centering
\includegraphics[width=0.5 \textwidth]{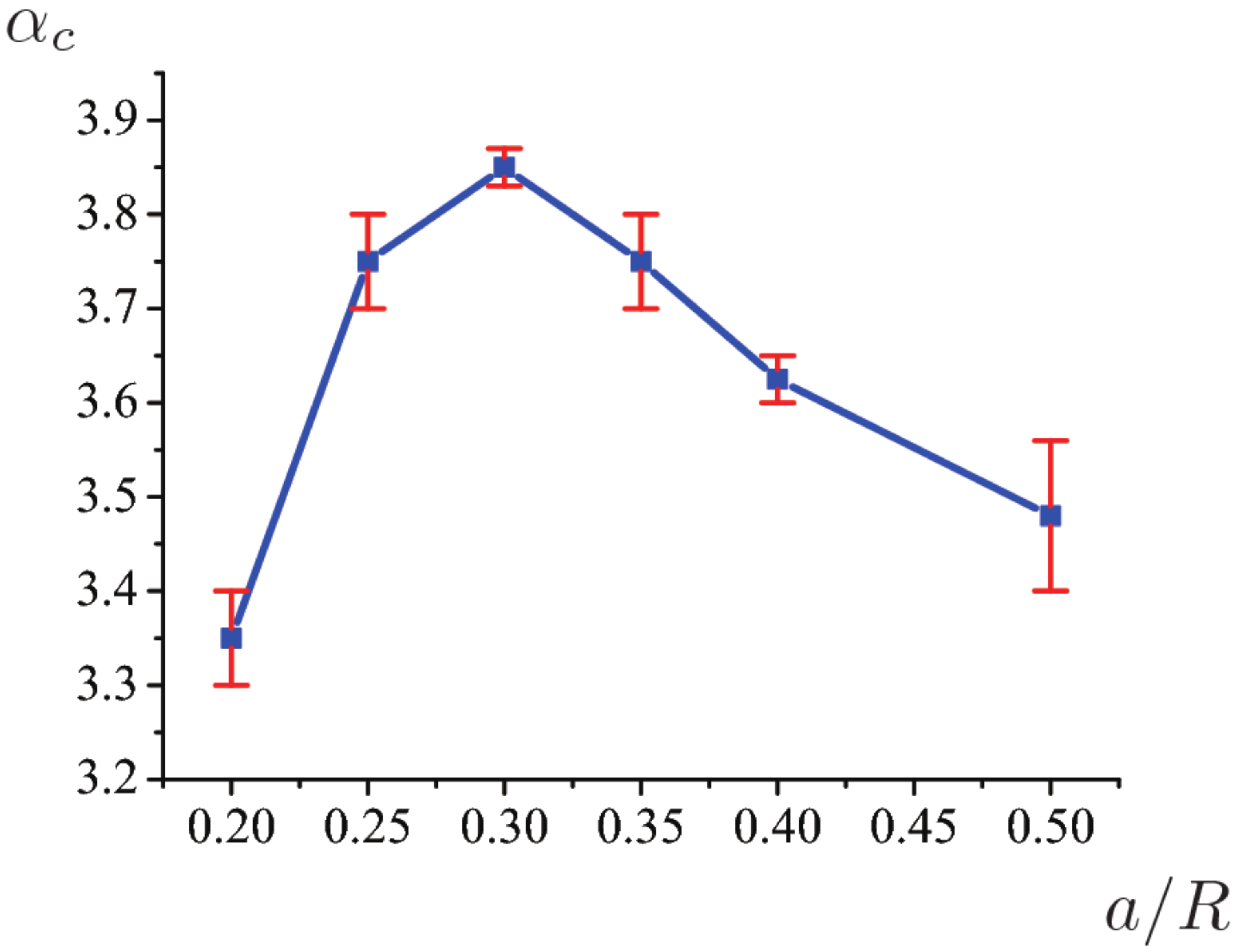}
\caption{Critical value of the dipole parameter, $\alpha_{c}$, for  stable swimming of the puller in the tube centre as a function of  the
confinement, $a/R$. The approximate values of $\alpha_{c}$ are denoted by the square symbols and the upper (resp.~lower) limit of the error bar corresponds to the asymptotically-stable swimming motion away from the centre (resp.~along the tube axis).}
\label{fig:crialpha}
\end{figure}

We further examine the coordinates of equilibrium points $\left(r_{CYL}, \xi_{CYL}\right)$ as a function of the  dipole strength, $\alpha$, in  figure~\ref{fig:pullerbranch}. For $\alpha$ below a critical value, $\alpha_{c}\approx 3.86$ for the confinement chosen here ($a/R=0.3$), the equilibrium point is $\left(r_{CYL}, \xi_{CYL}\right) = \left(0,0\right)$ denoted by the dashed blue line. For $\alpha>3.86$, the equilibrium point corresponds to swimming stably along a straight line with off-axis distance $r_{CYL}$ and orientation $\xi_{CYL}$, both of which grow with increasing $\alpha$. The relationship between confinement, $a/R$, and the critical value $\alpha_{c}$ is examined in figure~\ref{fig:crialpha}. Determining precisely the value of $\alpha_{c}$ is not possible due to the large computational cost so we report  approximate values, with  an upper (resp.~lower) limit of the error bar corresponding to the asymptotically-stable swimming motion near the wall (resp.~along the tube axis). The critical dipolar strength first increases with the confinement, reaching its maximum as $a/R \approx 0.3$,  before  decreasing.

\begin{figure}
   \centering
   \includegraphics[width=0.69 \textwidth]{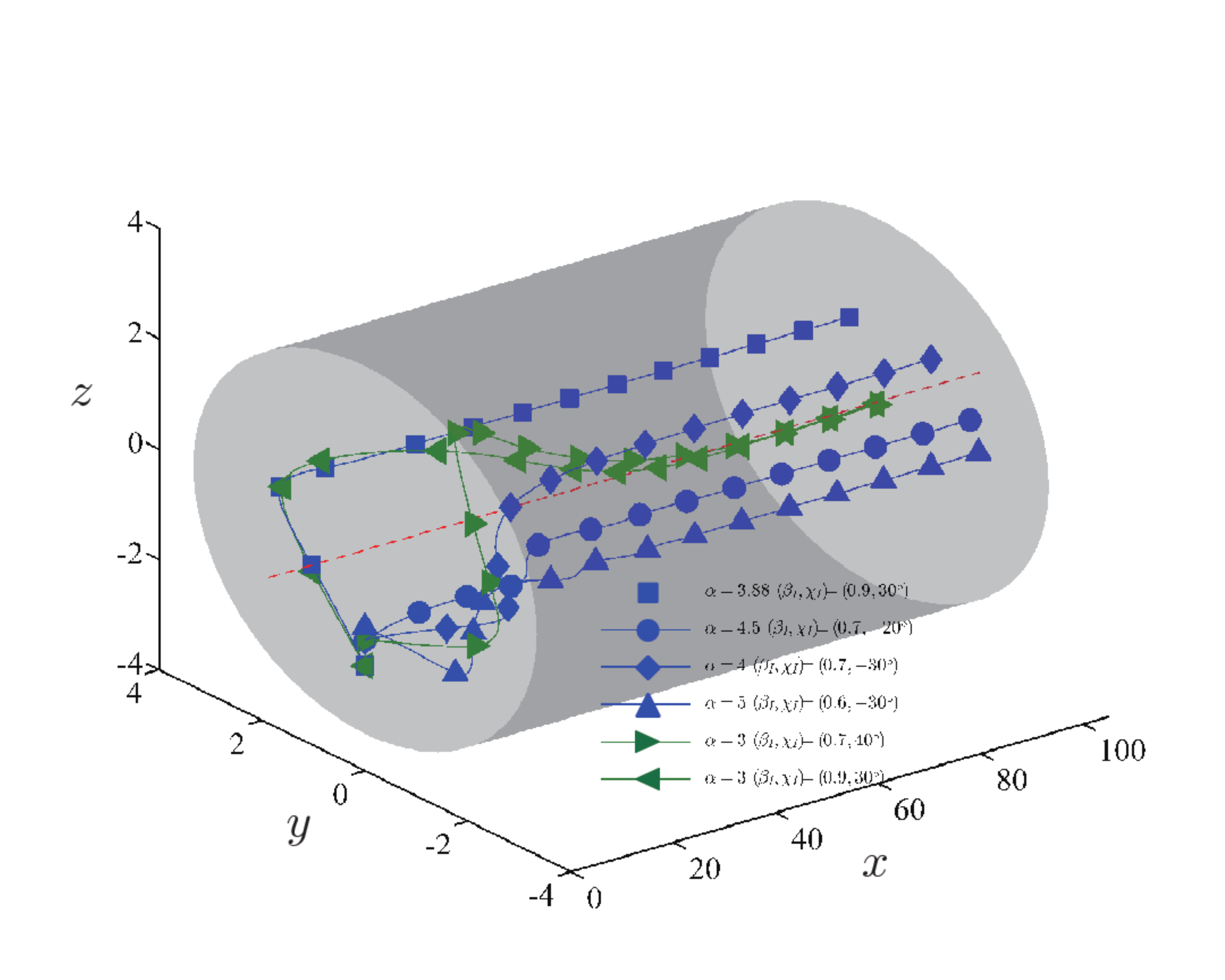}
   \caption{Three-dimensional trajectories of pullers in the tube with confinement $a/R=0.3$. The red dashed line indicates the tube axis. Different combinations of the dipole strength, $\alpha$,  initial position, $\beta_{I}$, and initial orientation,
$\chi_{I}$, are chosen. Green trajectories
correspond to pullers with one equilibrium point in the tube centre whereas  blue ones are for pullers with  equilibrium near the wall.}
   \label{fig:puller_tra_3D}
\end{figure}

By starting with different combinations of $\alpha$, $\beta_{I}$ and $\chi_{I}$, we obtain different three-dimensional trajectories for the puller. Some of these trajectories are illustrated in figure~\ref{fig:puller_tra_3D}. Results similar to the two-dimensional simulations are obtained. For  $\alpha$   below a critical value, pullers eventually swim along the tube axis indicating the equilibrium point $\left(r_{CYL},\xi_{CYL}\right) = \left(0,0\right)$. For larger values of $\alpha$, the equilibrium point corresponds to swimming motion with constant off-axis distance and orientation.  
Hydrodynamic interactions between the swimmer and the tube alone are responsible for such a passive stability, which could be of importance to guarantee, for example, robust steering of artificial micro-swimmers in capillary tubes without on-board sensing and control~\citep{yizhar_pre_stability}.

\begin{figure}
   \centering
   \includegraphics[width=0.3 \textwidth]{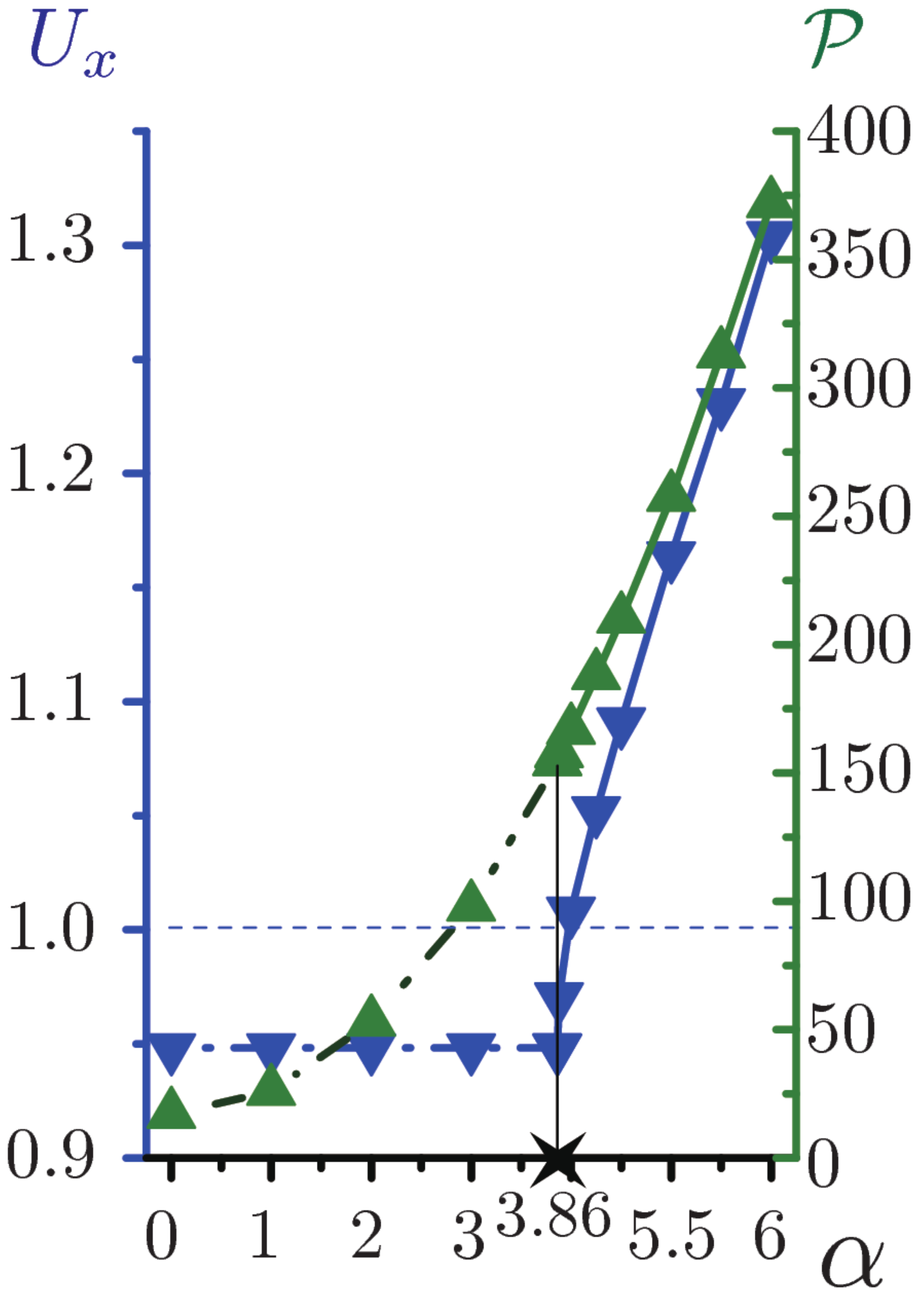}
   \caption{Swimming speed, $U_{x}$ (blue lower triangles), and power consumption, $\mathcal{P}$ (green upper triangles), of the puller as a function of the  dipole parameter, $\alpha$, both quantities being scaled by their corresponding values in free space (with $a/R=0.3$). The critical value $\alpha_c = 3.86$, shown by the cross, is the transition between stable swimming at the tube centre vs.~stable swimming near the tube walls. The blue dashed line indicates the swimming speed in free space.}
   \label{fig:pullBran_vel_pow}
\end{figure}

\begin{figure}
   \centering
   \includegraphics[width=0.64
\textwidth]{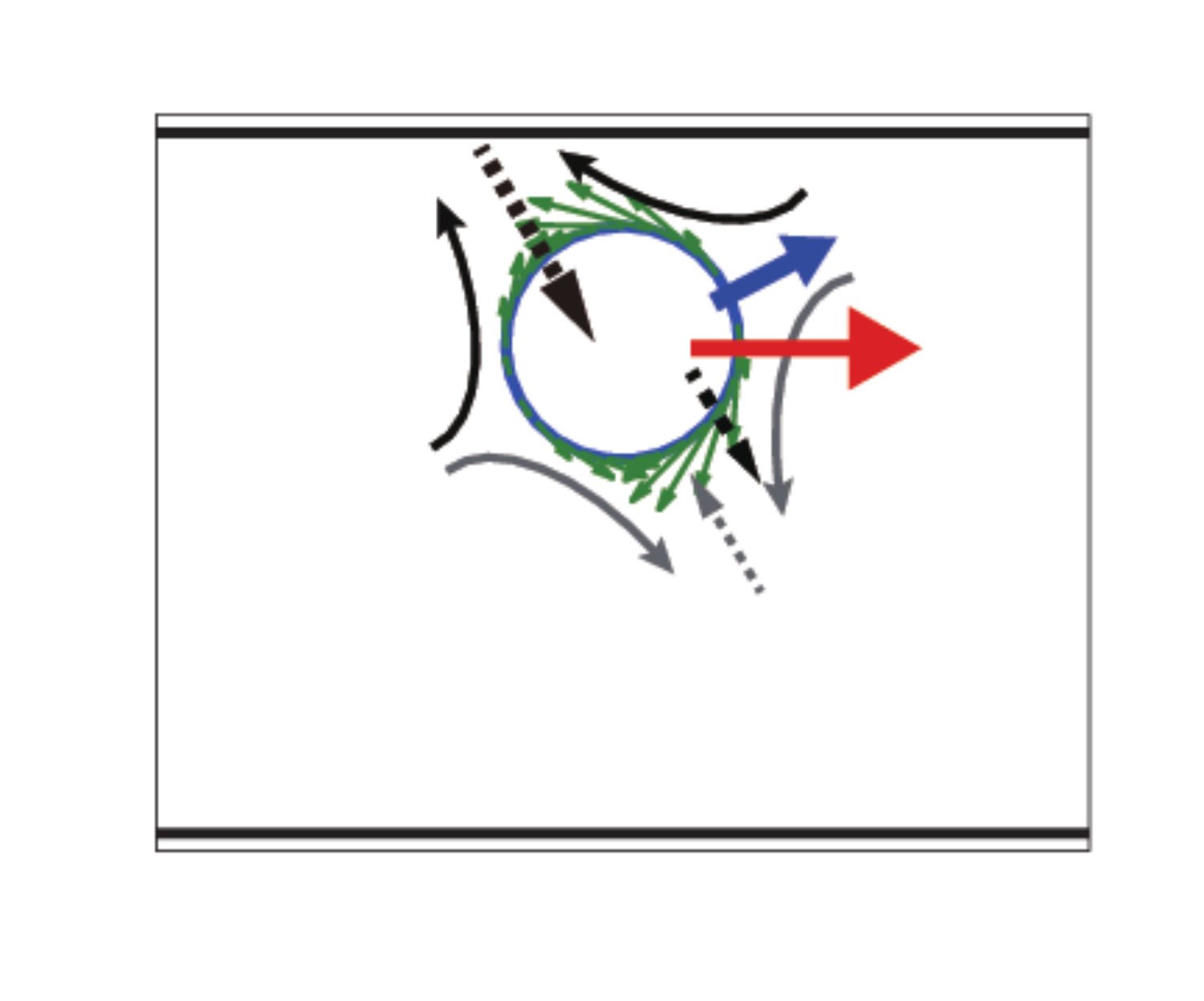}
   \put(-125,140){\small{$\mathbf{F}_{N}^{P}$}}
   \put(-118,120){\small{$\mathbf{F}_{R}$}}
   \put(-70,156){\small{$\hat{\mathbf{e}}$}}
   \put(-50,131){\small{$\mathbf{u}$}}
   \put(-90,70){\small{$\mathbf{F}_{F}^{P}$}}
   \caption{The orientation, $\hat{\mathbf{e}}$, of a puller swimming (red arrow) on the stable trajectory near the tube wall. Curved green arrows stand for the   flow imposed at the surface of the swimmer, black ($\mathbf{F}_{N}^{P}$) and grey ($\mathbf{F}_{F}^{P}$) dashed arrows for the hydrodynamic force while $\mathbf{F}_{R}$ is the repulsive force.}
   \label{fig:puller_near}
\end{figure}

We conclude this section by investigating in figure~\ref{fig:pullBran_vel_pow} the swimming speed of the puller along the stable trajectory and the dependency of its magnitude on the  dipole strength, $\alpha$. {In the case of confinement $a/R=0.3$}, the swimming speed $U_{x}$ is larger than that in free space as $\alpha$ is above a critical value (around $4$ here)  and it increases by about $16\%$ as $\alpha=5$. This is an example of swimming microorganisms taking propulsive advantage from near-wall hydrodynamics, as discussed in previous analytical studies~\citep{katz_sperm_bound, Felderhol_twoplane, Felderhol_pof_pipe}. In our case, as the squirmer is oriented into the wall, the direction of the wall-induced hydrodynamic force, $F_{R}$, resulting from flow being ejected on the side of the puller, is not normal to the wall but possesses a component in the swimming direction, as shown in figure~\ref{fig:puller_near}. This force contributes thus to an  additional propulsion and increases the swimming speed.

\subsection{The trajectory of a pusher inside the tube}\label{pusher}

We next address the spherical pusher squirmer, with a  negative  force dipole, $\alpha$. We find  that the motion of the pushers inside the tube is unstable. The trajectories of pushers confined in the $x-z$ plane ($\chi_{I}=0$) are plotted in figure~\ref{fig:pusher_branch2d} for different combinations of  dipole strength,  initial position, and initial orientation. The pushers always execute wavelike motions with decreasing wavelengths and increasing amplitude, eventually crashing into the walls.  
{Pushers and pullers display therefore very different swimming behaviours, a difference which stems from the opposite  front-back asymmetry of  the force dipole.

\begin{figure}
   \centering
   \includegraphics[width=0.6 \textwidth]{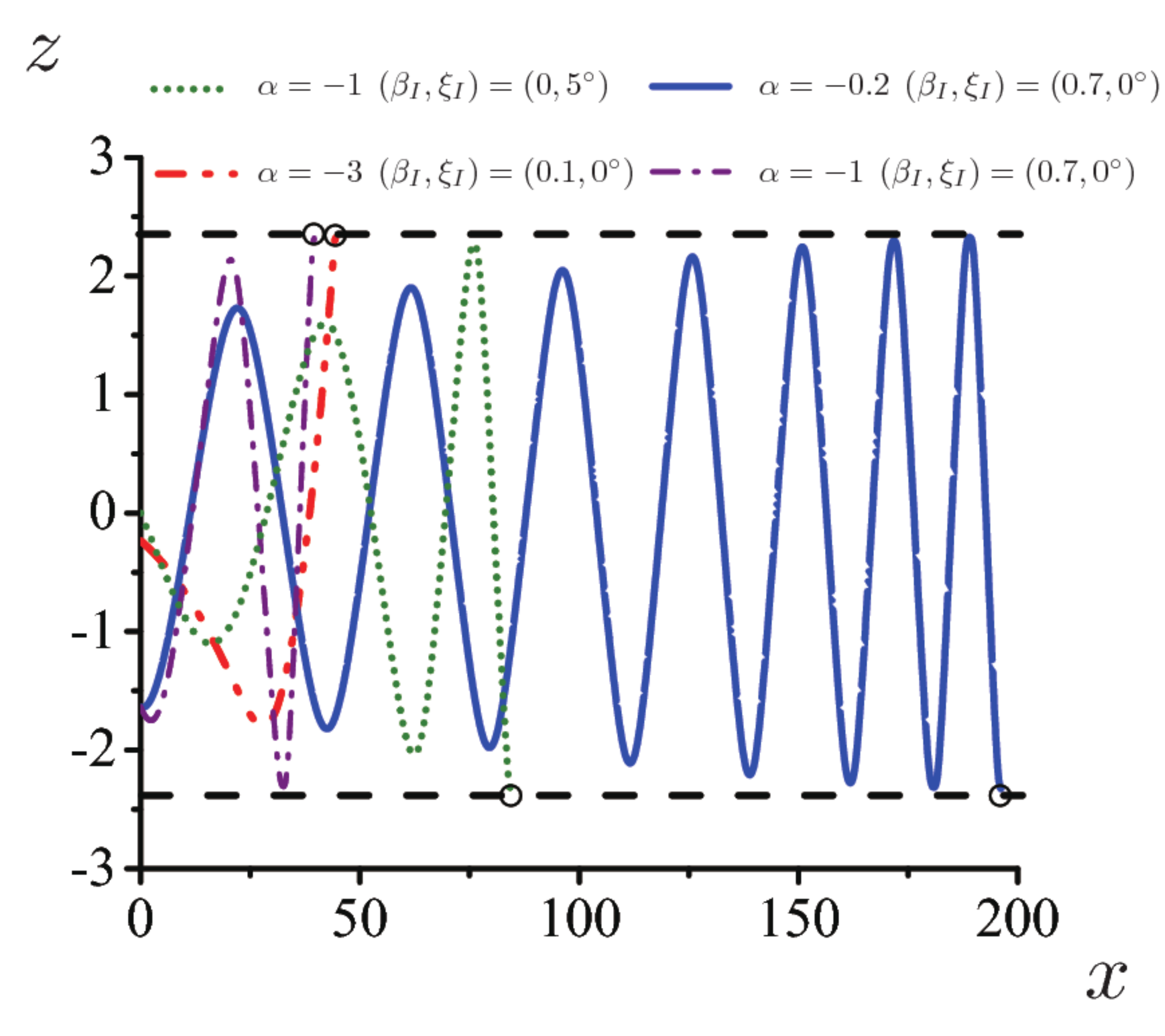}
   \caption{Two-dimensional trajectories of pushers for different combinations of $\alpha$, $\beta_{I}$, and $\xi_{I}$. The black circles indicate the moment  the swimmers make contact with the wall.}
   \label{fig:pusher_branch2d}
\end{figure}

\subsection{Squirmers with normal surface velocity}\label{jet}

For the sake of completeness, we investigate in this section the dynamics of squirmers in the tube in the case where the squirming motion is induced by normal (instead of tangential) surface velocity, modelled as
\begin{equation}
 \mathbf{u}_{SN} ({\bf r})=\sum_{n=0} \frac{2}{n(n+1)}A_n
  P_{n}\left(\frac{\hat{\bf{e}} \cdot \mathbf{r}}{r}\right),
\end{equation}
where $A_n$ is the $n$th mode of the normal squirming velocity~\citep{Blake1971a}. In
free space, the swimming velocity is $U_{SN}^{F}=-{A_{1}}/{3}$~\citep{Blake1971a}. For simplicity, we
only consider the instantaneous kinematics of a squirmer with $A_{1} = -1$ and $A_{n\neq1}=0$, corresponding thus to $U_{SN}^{F}={1}/{3}$. The swimmer is located at $\left(0,0,-\beta\left(R-a\right)\right)$, and is oriented in the positive $\mathbf{x}$ direction. We plot the axial velocity component, $U_{x}$, (scaled by $U_{SN}^{F}$) together with the rotational velocity, $\Omega_{y}$, in figure~\ref{fig:spd_normal}. Both $U_{x}$ and $\Omega_{y}$ are seen to increase monotonically with the confinement and eccentricity. This is in agreement with past mathematical analysis stating that microorganisms utilising transverse surface displacement speed up when  swimming near  walls~\citep{katz_sperm_bound}, between two walls~\citep{Felderhol_twoplane}, or inside a tube~\citep{Felderhol_pof_pipe}.

This increase (resp.~decrease) of swimming speed  in the tube of a squirmer with normal (resp.~tangential)  surface deformation can be related to the problem of  micro-scale locomotion in polymeric solutions. It is well known that actuated biological flagella generate drag-based thrust  due to larger resistance to normal than to tangential motion \citep{LaugaSwim}. When swimming in polymer solutions, flagella undergoing  motion normal to its shape push  directly onto the neighbouring polymer network, whereas tangential motion barely perturb these micro obstacles~\citep{berg79_cockscrew, polynetworkliketube,polySpiro,leshansky09}. In this case, the drag force  increases more in the normal direction than in the tangential, resulting in larger swimming speeds~\citep{berg79_cockscrew,polynetworkliketube,polySpiro,leshansky09,LiuHelixPnas}.  Likewise, it was shown for a spherical squirmer that polymeric structures in the fluid always decrease the swimming speed in case of tangential surface deformation~\citep{leshansky09,lailai-pre,laipof1} but increase for normal deformation~\citep{leshansky09}.  The increase of swimming speed observed here in the case of a squirmer with normal surface deformation can similarly be attributed to the flow directly onto the tube wall.

\begin{figure}
   \centering
   \includegraphics[width=0.9 \textwidth]{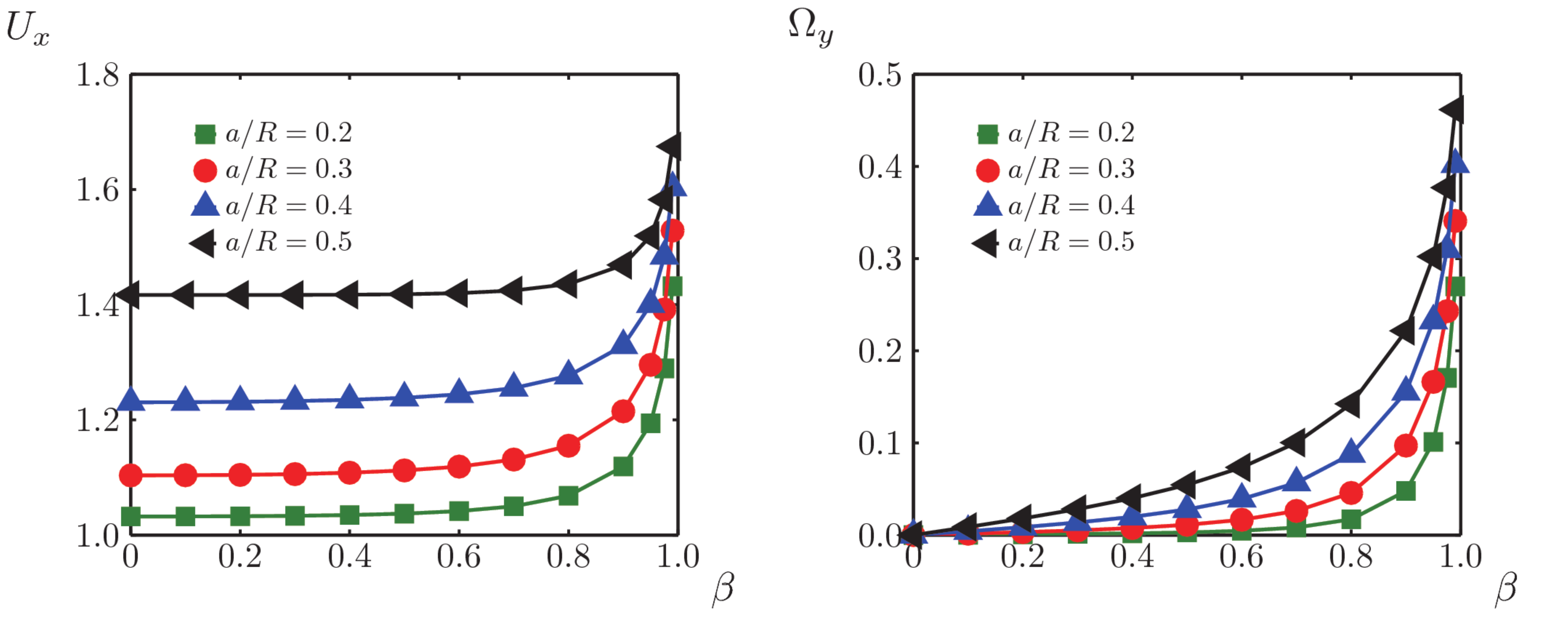}
   \caption{Swimming velocity in the axial direction, $U_{x}$, and rotational velocity, $\Omega_{y}$, of the squirmer with normal surface deformation with modes $A_{n}=-\delta_{n1}$; $U_{x}$ is scaled by the swimming speed in free space, $U_{SN}^{F}$. The squirmer is located at $\left(0,0,-\beta\left(R-a\right)\right)$ and oriented parallel to the axis. Different values of $a/R$ are reported with maximum value of $\beta=0.99$.}
\label{fig:spd_normal}
\end{figure}

The value of rotational velocity, $\Omega_{y}$, shown in figure~\ref{fig:spd_normal} shows however that the squirmer rotates into the nearest wall,  thus getting eventually trapped there. In order to avoid being trapped while at the same time taking advantage of the wall-induced enhanced propulsion, ideally swimmers should thus use a combination of tangential and normal deformation. 

{Interestingly, a superposition of the neutral squirming mode
($B_n = \delta_{n1}$, see \S\ref{mathmodel}) with the first normal squirming mode ($A_n = -\delta_{n1}$) results in a special
swimmer able to move without creating any disturbance in the surrounding 
fluid, characterised by a uniform squirming velocity of $-1$ everywhere on the body (in the co-moving
frame), no body rotation, and a swimming speed equal to $1$. This remains true regardless of the degree
of confinement as confirmed by our numerical simulations.}

\section{Swimming inside a curved tube}\label{bent}

\begin{figure}
   \centering
   \includegraphics[width=0.75 \textwidth]{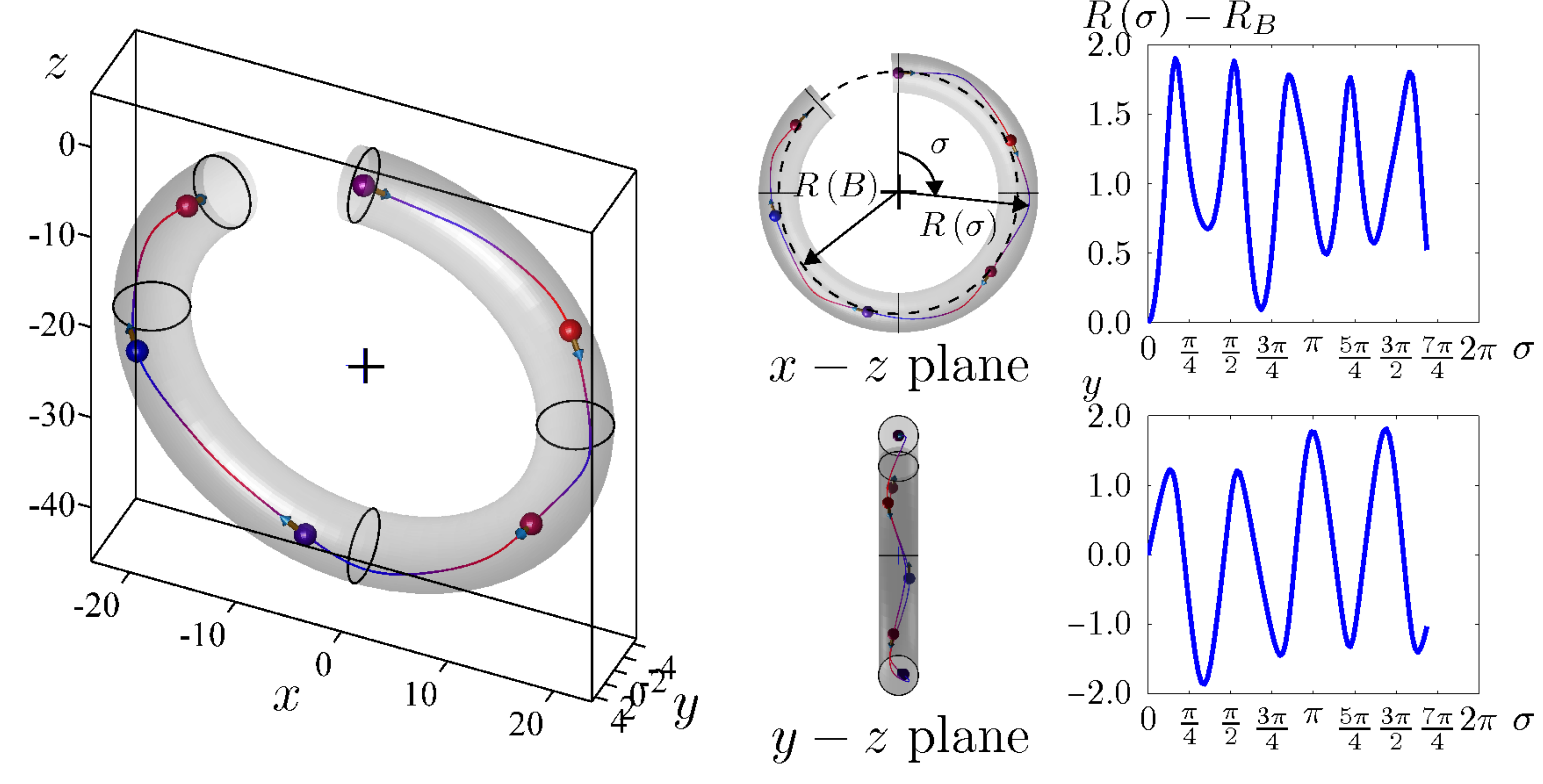}
   \caption{Three-dimensional swimming of the neutral squirmer inside a torus-like curved tube. The dashed line indicates the circular axis of the torus with its radius, $R_{B}=20a$,  $\sigma$ is the azimuthal position of the squirmer, and $R\left(\sigma\right)$ is the distance between the squirmer and the centre of the baseline circle. The wavelike motions for $R\left(\sigma\right)-R_{B}$ and $y(\sigma)$ are shown on the right.}
   \label{fig:bent_neutral}
\end{figure}

In this final section, we investigate the squirmer motion inside a curved tube that is a part of a torus. The axis of the torus is a circle on the plane $y=0$ with its radius $R_{B}=20a$. Trajectories of a neutral squirmer and a puller with the  dipole strength $\alpha=1$ are shown in Figs.~\ref{fig:bent_neutral} and \ref{fig:bent_puller_al1} respectively. In both cases, the trajectory is displayed in both the $x-z$ and $y-z$ planes. The motion in the radial direction, represented by $R\left(\sigma\right)-R_{B}$, is plotted as a function of the azimuthal position of the swimmer, $\sigma$, where $R\left(\sigma\right)$ is the distance between the cell and the centre of the circle. In both cases, the dynamics of swimmers initially starting aligned with the tube axis is wavelike. For the neutral squirmer, the wavelength and wave magnitude approach a constant value $\sigma>\pi$ (figure~\ref{fig:bent_neutral}, right), indicating marginal stability of the motion. In contrast, for the puller, decaying waves are observed (figure~\ref{fig:bent_puller_al1}, right), indicating passive asymptotic stability. As in the straight-tube case, pushers are unstable and crash into walls in finite time.

\begin{figure}
   \centering
   \includegraphics[width=0.75 \textwidth]{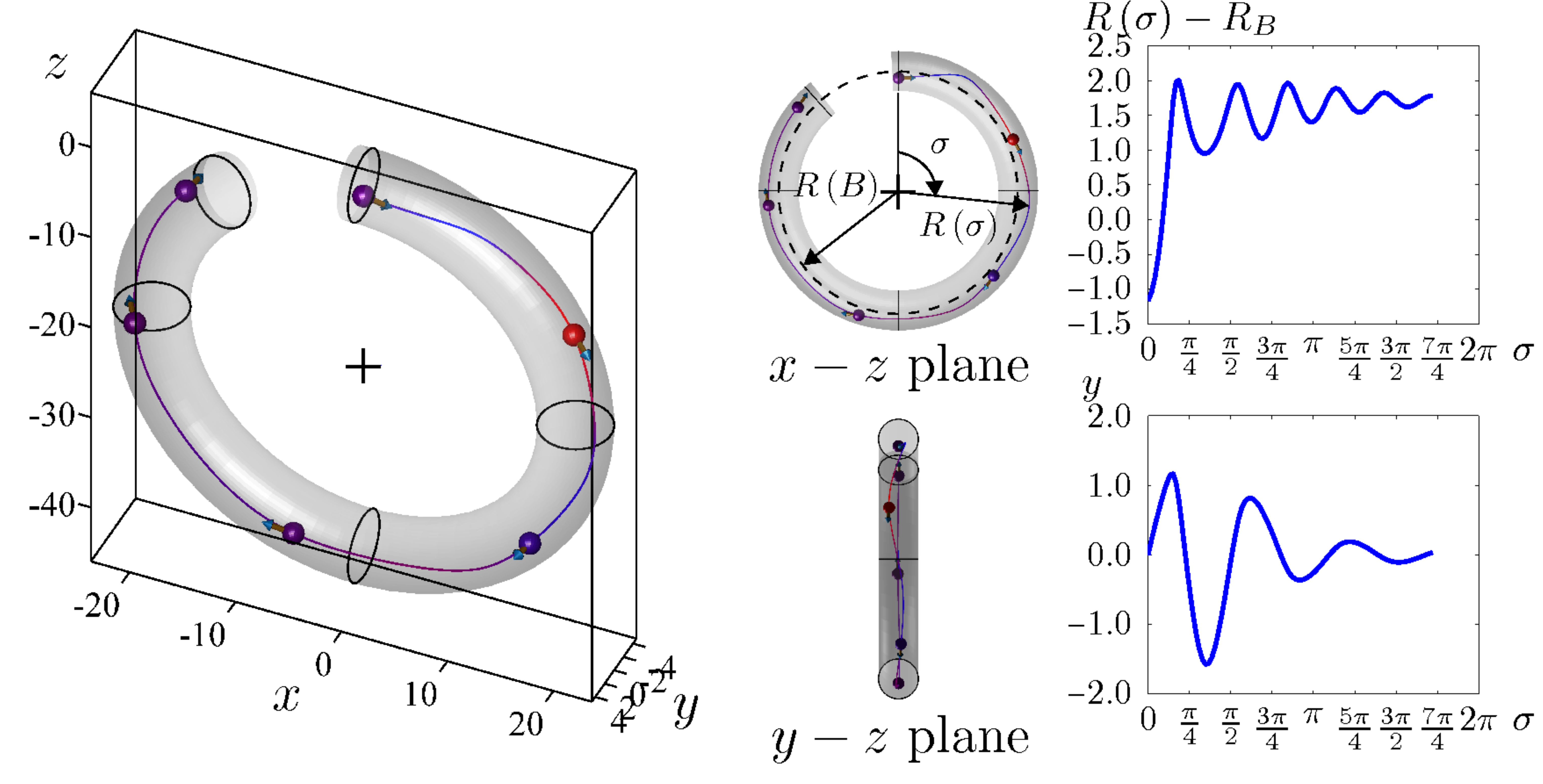}
   \caption{Same as in figure~\ref{fig:bent_neutral} but for a puller with $\alpha=1$.}
   \label{fig:bent_puller_al1}
\end{figure}


\section{Conclusion and outlook}

In this paper, a Boundary Element Method code was developed, validated, and used to present  computations for the locomotion of model  ciliates inside straight and curved capillary tubes.  We used the spherical squirmer as our model microorganism and studied the effect of confinement on the kinematics, energetics, and trajectories of the cell. We also investigated the stability of the swimming motion of squirmers with different gaits (neutral, pusher, puller). 

We found that tube confinement and near-wall swimming  always decrease the swimming speed of a squirmer with tangential surface deformation for swimming parallel to the tube axis. In contrast, a swimmer with normal surface deformation improves its swimming speed by directly pushing against the surrounding tube wall. In both cases however, tube confinement and near-wall swimming always lead to additional viscous dissipation, thus increasing the power consumption.

Focusing on swimming with tangential forcing, we then studied in detail the dynamics of  neutral, puller, and pusher squirmers inside a straight tube. For a neutral squirmer,  swimming motion on the tube axis is  marginally stable and generically displays three-dimensional  helical trajectories as previously observed experimentally  for \textit{Paramecium} cells. Importantly, these helical trajectories arise purely from hydrodynamic interactions with the boundaries of the tube.

In the case of puller swimmers, their trajectories   are wavelike with  decreasing amplitude and increasing wavelength, eventually leading to stable swimming parallel with the tube axis with their bodies slightly oriented toward the nearest wall. The locations for these stable trajectories depend on the strength of the force dipole, $\alpha$. Swimmers with weak dipoles (small $\alpha$) swim in the centre of the tube while those with  strong dipoles (large $\alpha$) swim near the walls. The stable orientation of the swimmers makes an non-zero contribution of the wall-induced hydrodynamic forces in the direction of locomotion, thus leading to an increase of the swimming speed (although accompanied by an increase of the rate of viscous dissipation). In contrast, pushers are always unstable and crash into the walls of the tube in finite time. Similar results are observed for locomotion inside a curved tube.

We envision that our study and general methodology could be useful in two specific cases. First, our results could help shed light on and guide the future design and maneuverability of artificial small-scale swimmers inside small  tubes and conduits. Second, the computational method could be extended to more complex, and biologically-relevant, geometries, to study for example the locomotion of flagellated bacteria or algae into confined geometries, as well as their hydrodynamic interactions with relevant background flows. It would be also interesting to relax some of our assumptions in future work, and address the role of swimmer geometry on their stability (we only considered the case of spherical swimmers in our paper)  and quantify the role of noise and fluctuations on the asymptotic dynamics obtained here.

\section*{Acknowledgements}
We thank Prof. Takuji Ishikawa for useful discussions.
Funding  by VR (the Swedish Research Council) and the National Science Foundation (grant CBET-0746285 to E.L.) is gratefully acknowledged.
Computer time provided by SNIC (Swedish National Infrastructure for Computing) is also acknowledged.

\bibliographystyle{jfm}
\bibliography{liczhu}

\end{document}